\documentclass[preprint,12pt]{elsarticle}
\usepackage{graphicx}
\usepackage[outdir=./]{epstopdf}
\usepackage{booktabs} 
\usepackage{url}
\usepackage{amsmath}
\usepackage{tabu}
\usepackage{paralist}
\usepackage[linewidth=1pt,nobreak=true]{mdframed}
\usepackage{lipsum}
\usepackage{subcaption}
\usepackage{color, colortbl}
\usepackage{xcolor}
\usepackage{makecell}
\usepackage{ltablex}
\usepackage{booktabs,multicol,multirow,graphicx,tabularx,ragged2e,dcolumn}
\definecolor{LightCyan}{rgb}{0.88,1,1}
\definecolor{Gray}{gray}{0.9}
\usepackage{xspace}
\usepackage{enumitem}
\usepackage{natbib}
\usepackage{lscape}
\usepackage{caption} 
\newcommand*{\eg}{\emph{e.g.},\@\xspace}
\newcommand*{\ie}{\emph{i.e.},\@\xspace}
\newcommand*{\aka}{\emph{a.k.a.}\@\xspace}

\newcommand*{\vsepfbox}[1]{%
	\begingroup
	\sbox0{\fbox{#1}}%
	\setlength{\fboxrule}{0pt}%
	\mbox{\kern-\fboxsep\fbox{\unhbox0}\kern-\fboxsep}%
	\endgroup
}
\renewcommand{\arraystretch}{1.3}

\makeatletter
\newcommand*{\etc}{%
	\@ifnextchar{.}%
	{etc}%
	{etc.\@\xspace}%
}
\newcommand\tandoori[0]{\textsc{Sniffer}}
\newcommand\paprika[0]{\textsc{Paprika}}

\begin{document}
	\title{Android Code Smells:\\ \em From~Introduction to~Refactoring}
	
	\author[Affil1]{Sarra Habchi\corref{cor1}\fnref{fn1}}
	\ead{sarra.habchi@uni.lu}
	\cortext[cor1]{Corresponding author}

	\author[Affil2]{Naouel Moha\fnref{fn2}}
	\ead{moha.naouel@uqam.ca}
	
	\author[Affil3]{Romain Rouvoy\fnref{fn2}}
	\ead{romain.rouvoy@inria.fr}
	\address[Affil1]{University Of Luxembourg}
	\address[Affil2]{Université du Québec À Montréal}
	\address[Affil3]{University of Lille }
	
	\begin{abstract}
Object-oriented code smells are well-known concepts in software engineering that refer to bad design and development practices commonly observed in software systems. 
With the emergence of mobile apps, new classes of code smells have been identified by the research community as mobile-specific code smells. 
These code smells are presented as symptoms of important performance issues or bottlenecks.
Despite the multiple empirical studies about these new code smells, their diffuseness and evolution along change histories remains unclear. 

We present in this article a large-scale empirical study that inspects the introduction, evolution, and removal of Android code smells. 
This study relies on data extracted from $324$ apps, a manual analysis of $561$ smell-removing commits, and discussions with $25$ Android developers. 
Our findings reveal that the high diffuseness of mobile-specific code smells is not a result of releasing pressure. 
We also found that the removal of these code smells is generally a side effect of maintenance activities as developers do not refactor smell instances even when they are aware of them. 
\end{abstract}

	\maketitle

	\section{Introduction}\label{sec:introduction}
Mobile apps have established themselves as mainstream software systems deployed at scale.
Over the last few years, they successfully invaded the software market and retained the interest of end-users.
While mobile apps generally rely on the same software development basics as classical software systems, they also manifest some particularities as they run on embedded devices with key performance constraints.
This specificity sets the bar high for mobile apps in the sense that they are expected to remain fluid and efficient while continuously performing complex tasks.
As a consequence, development practices that do not satisfy these requirements were qualified as \emph{code smells}.
In particular, the study of Reimann~\emph{et~al.}~\cite{Reimann2014} proposed a catalogue of Android-specific code smells that violate performance guidelines.
These code smells originate from the good and bad practices presented in the official documentation or by developers reporting their experience on blogs.
For example, the code smell \emph{No Low Memory Resolver} describes non-compliance with the official recommendation of implementing memory resolvers inside activities.

The research community extended this catalogue and studied different aspects of mobile-specific code smells.
The performance impact of these code smells was inspected in multiple studies~\cite{carette2017investigating,hecht2016empirical,palomba2019impact}, showing that they can hinder app performance and increase energy consumption.
Research works also proposed automated solutions for detecting mobile code smells, like \paprika{} and \textsc{aDoctor}~\cite{hecht2015tracking,palomba2017lightweight}.
On the empirical side, our previous studies assessed the key role played by developers in the accrual of mobile code smells~\cite{habchi2018adopting,habchi2019rise} and quantified their survival in the change history.
This allowed us to build an initial understanding of what are mobile code smells and how do they hinder app quality.
However, as research remains young in this field, we still lack knowledge about various aspects of code smells:

\paragraph{\textbf{Lack of extent analysis}} 
Studies proposing code smell detection tools quantified code smell instances in Android and iOS apps~\cite{habchi2017code,hecht2015tracking,palomba2017lightweight}.
However, none of these studies analysed and compared the diffuseness of code smells of different types.
This comparison is important to distinguish the most common code smells and prioritize their detection and refactoring in future studies. 
Diffuseness analysis is also important to precisely assess code smell prevalence.
In particular, a code smell may seem very frequent only because its host entity is very common in source code. 
Diffuseness analysis alleviates this by measuring precisely how frequent is the code smell considering its occurrence chances and its host entity.

\paragraph{\textbf{Lack of release analysis}}
Some studies~\cite{habchi2019rise,habchi2019survival} leveraged the change history of mobile apps to better understand code smells.
Specifically, our previous work~\cite{habchi2019survival} evaluated the impact of releases on code~smell survival.
However, this study did not assess the impact of releases on the introduction and removal of code smells.
Releases are usually considered as a factor that favours code smells and technical debt in general, since they push developers to code rapidly and meet deadlines regardless of quality constraints~\cite{tom2013exploration}.
Besides, mobile apps are known for having more frequent releases and updates~\cite{mcilroy2016fresh}, which may contribute to the prevalence of mobile code smells.
Considering these potential factors, it is important to analyse the impact of releases on the presence of mobile code smells.

\paragraph{\textbf{Lack of qualitative analysis}}
In a previous work~\cite{habchi2018adopting}, we interviewed developers to understand their usage of linters to anticipate performance bottlenecks in mobile apps.
This study gave insights about the adequacy of static analysers as a solution for mobile code smells.
Nonetheless, other facets of these code smells still require qualitative investigation.
In particular, we lack knowledge about how do developers remove mobile code smells from the source code.
This knowledge is important to:
\begin{compactitem}
	\item Assess developers' awareness of mobile code smells;
	\item Check whether developers refactor these code smells intentionally or not;
	\item Learn removal techniques from developers and aliment future studies about code smell refactoring.
\end{compactitem}

\bigskip
In this article, we address these lacks by answering the following research questions:
\begin{compactitem}
	\item \textsc{\textbf{RQ\,1:}} How frequent and diffuse are mobile code smell introductions?
	\item \textsc{\textbf{RQ\,2:}} How do releases impact introductions and removals of mobile code smells?
	\item \textsc{\textbf{RQ\,3:}} How do developers remove mobile code smells?
	\item \textsc{\textbf{RQ\,4:}} Do developers refactor mobile code smells?
\end{compactitem}

\bigskip
To answer these questions, we build on the artifacts of our previous works~\cite{habchi2019rise,habchi2019survival} to perform an empirical study where we leverage both quantitative and qualitative analyses to inspect introductions and removals of $8$ types of Android code smells.
Specifically, we analyse the evolution of $180k$ code smell instances to answer \textsc{RQ\,1} and \textsc{RQ\,2}.
Then, we manually explore $561$ code smell removals to answer \textsc{RQ\,3} and finally we interview $25$ smell-removing developers to answer \textsc{RQ\,4}.
The results of this study show that:
\begin{compactenum}
	\item Regarding frequency and diffuseness, there is an important discrepancy between code smell types.
	 \emph{No~Low~Memory~Resolver} and \emph{Leaking~Inner~Class} are the most diffuse by affecting more than $80\,\%$ of the activities and inner classes, respectively.
	\item Releases do not have an impact on the introductions and removals of code smells in open-source Android apps.
	\item $79\%$ of code smell instances are removed through the change history.
	However, these removals are mostly caused by large source code removals that do not mention refactoring.
	Also, only $19\%$ of developers who authored these removals confirmed that their actions were intentional refactoring.
	\item Developers who are aware of Android code smells do not necessarily refactor them.
	The code smell \emph{Init OnDraw} was recognized by $64\%$ of the participants, but only $12\%$ of them refactored it.
	\item Developers who intentionally refactor code smells affirm that their actions were driven and assisted by built-in code analysis tools.
	\item Developers who did not refactor Android code smells doubted their performance impact and the usefulness of their refactoring.
	Some developers also preferred to handle performance issues when they arise instead of anticipating them.
\end{compactenum}

\bigskip
This study provides a comprehensible replication package~\cite{CompanionArtifacts}, which includes the used tools and data analysis scripts, the extracted data, and the results of the qualitative analysis.

\bigskip
The remainder of this article is organized as follows.
Section~\ref{sec:study_design} explains the study design, while Section~\ref{sec:results} reports on the results.
Section~\ref{sec:discussion} interprets and discusses these results, and Section~\ref{sec:threats} exposes the threats to validity.
Finally, Section~\ref{sec:related_work} analyses related works, and Section~\ref{sec:conclusion} concludes with our main findings.

 	\section{Study Design}\label{sec:study_design}
To perform this study, we relied on the artifacts that we built in our previous works about mobile code~smells.
In particular, we leveraged the dataset of code~smell history~\cite{habchi2019rise,habchi2019survival} to collect the necessary data for this study.
Then, we followed different approaches to analyse this data and answer our research questions.

\subsection{Dataset}
In previous works, we created a dataset containing the history of mobile-specific code~smells.
This dataset was built by running \tandoori{}~\cite{snifferSource} on a set of Android apps and tracking $8$ mobile code~smells.
For self-containment purposes, we present in this section (i) \tandoori{}, (ii) the $8$ code~smells, and (iii) the contents of this dataset.

\subsubsection{Sniffer}
\tandoori{} is an open-source~\cite{snifferSource} toolkit that tracks the full history of Android-specific code~smells.
It tackles many issues raised by the Git mining community by tracking branches and detecting renaming~\cite{kovalenko2018mining}.
\tandoori{} builds the code~smell history by following a three-step process.
First, from the repository of the app understudy, it extracts the commits and other necessary metadata like branches, releases, and commit authors.
In the second step, it analyses the source code of each commit separately to detect code~smell instances.
Finally, based on the code~smell instances and the repository metadata, it tracks the history of each smell and records it in the output database. 

The performance of \tandoori{} was manually validated using $384$ commits randomly sampled from open-source Android apps.
This validation showed that it can detect code~smell introductions with F1-score of $0.97$ and code~smell removals with a score of $0.96$.


\subsubsection{code~smells}
The dataset covers all the $8$ types of Android-specific code~smells that are detectable by \tandoori{}.
These code~smells are performance-oriented and they originate from the catalogues of Reimann~\emph{et~al.}~\cite{Reimann2014} and Hecht~\emph{et~al.}~\cite{hecht2017detection,hecht2015tracking}.
Unlike other Android code~smells, these $8$ smells are objective---\ie they either exist in the code or not, and cannot be introduced or removed gradually.
Hence, their introduction and removal can be attributed to specific commits without confusion.
Table~\ref{table:code_smells} presents these code~smells with a highlight on source code entities in which they can appear.
We also mention the performance resource impacted by each code~smell.

\begin{tabularx}{\columnwidth}{X}
	\hline
	\hline
	\textbf{Leaking~Inner~Class (\textit{LIC})}:
	in Android, anonymous and non-static inner classes hold a reference of the containing class. This can prevent the garbage collector from freeing the memory space of the outer class even when it is not used anymore, and thus causing memory leaks~\cite{LintCheck,Reimann2014}.\\
	\textbf{Entity}: Inner class.\\
	\textbf{Impact}: Memory.\\
	\hline
	
	\textbf{Member~Ignoring~Method (\textit{MIM})}:
	this smell occurs when a method that is not a constructor and does not access non-static attributes is not static. As the invocation of static methods is 15\%--20\% faster than dynamic invocations, the framework recommends making these methods static~\cite{hecht2015tracking}.\\
	\textbf{Entity}: Method.\\
	\textbf{Impact}: CPU.\\
	\hline

	\textbf{No~Low~Memory~Resolver (\textit{NLMR})}:
	this code~smell occurs when an \texttt{Activity} does not implement the method \texttt{onLowMemory()}. This method is called by the operating system when running low on memory in order to free allocated and unused memory spaces. If it is not implemented, the operating system may kill the process~\cite{Reimann2014}.\\
	\textbf{Entity}: Activity.\\
	\textbf{Impact}: Memory.\\
	\hline
	
	\textbf{Hashmap~Usage (\textit{HMU})}:
	the usage of \texttt{HashMap} is inadvisable when managing small sets in Android. Using HashMaps entails the auto-boxing process where primitive types are converted into generic objects. The issue is that generic objects are much larger than primitive types, 16 and 4 bytes, respectively. Therefore, the framework recommends using the \texttt{SparseArray} data structure that is more memory-efficient~\cite{LintCheck,Reimann2014}.\\
	\textbf{Entity}: Method.\\
	\textbf{Impact}: Memory.\\
	\hline
	
	\textbf{UI~Overdraw (\textit{UIO})}:
	a UI Overdraw is a situation where a pixel of the screen is drawn many times in the same frame. This happens when the UI design consists of unneeded overlapping layers, \eg hidden backgrounds. To avoid such situations, the \texttt{canvas.quickreject()} API should be used to define the view boundaries that are drawable~\cite{LintCheck,Reimann2014}.\\
	\textbf{Entity}: View.\\
	\textbf{Impact}: GPU.\\
	\hline
	
	\textbf{Unsupported~Hardware~Acceleration (\textit{UHA})}:
	in Android, most of the drawing operations are executed in the GPU. Rare drawing operations that are executed in the CPU, \eg \texttt{drawPath} method in \texttt{android.graphics.Canvas}, should be avoided to reduce CPU load~\cite{hecht2017detection,IWR}.\\
	\textbf{Entity}: Method.\\
	\textbf{Impact}: CPU.\\
	\hline
	
	\textbf{Init~OnDraw (\textit{IOD})}:
	\aka DrawAllocation, this occurs when allocations are made inside \texttt{onDraw()} routines. The \texttt{onDraw()} methods are responsible for drawing \texttt{Views} and they are invoked 60\,times per second. Therefore, allocations (\emph{init}) should be avoided inside them in order to avoid memory churn~\cite{LintCheck}.\\
	\textbf{Entity}: View.\\
	\textbf{Impact}: Memory.\\
	\hline
	
	\textbf{Unsuited~LRU~Cache~Size (\textit{UCS})}:
	in Android, a cache can be used to store frequently used objects with the \emph{Least Recently Used} (LRU) API. The code~smell occurs when the LRU is initialized without checking the available memory via the \texttt{getMemoryClass()} method. The available memory may vary considerably according to the device so it is necessary to adapt the cache size to the available memory~\cite{hecht2017detection,UCS}.\\
	\textbf{Entity}: Method.\\
	\textbf{Impact}: Memory.\\
	\hline
	\caption{Studied code~smells.}
	\label{table:code_smells}  
\end{tabularx}


\subsubsection{Content}
Running \tandoori{} on a set of $324$ open-source Android apps resulted in a dataset with the history of all code~smell instances that appeared in these apps.
Table~\ref{table:database} summarizes the contents of this dataset.

\begin{table}[!htbp]
	\centering
	\caption{Content of the dataset.}
	\label{table:database}
	\resizebox{\columnwidth}{!}{%
	\begin{tabular}{ccccccc} 
		\hline
		\textit{\textbf{Apps}} & \textit{\textbf{Commits}} & \textit{\textbf{Files}} & \textit{\textbf{Smell Instances}}&  \textit{\textbf{Developers}} &\textit{\textbf{Releases}}& \textbf{\textit{Branches}} \\ \hline
		 $324$ & $255,798$ &         $1,455,617$                 &              $180,013$      &   $5,104$ & $11,118$ &       $21,210$            \\ \hline
	\end{tabular}
	}	
\end{table}

\subsection{Data Analysis} 
In this subsection, we describe our approach for analysing the collected data to answer our research questions.
Table~\ref{table:metrics} reports on the list of metrics that we defined for this purpose.

\begin{table}[!htbp]
	\centering
	\caption{Study metrics.}
	\label{table:metrics}
	\bgroup
	\small
	\begin{tabularx}{\linewidth}{llX}
		\hline
		& \emph{\textbf{Metric}}  & \emph{\textbf{Description}}\\
		\hline
		\multirow{13}{*}{\rotatebox[origin=c]{90}{\textbf{\emph{code~smell type}}}}
		& \textsf{\#introductions} & The number of instances introduced in the dataset.\\\cline{2-3}
		& \textsf{\%affected-apps} & The percentage of apps affected by the code~smell.\\\cline{2-3}
		& \textsf{\%diffuseness}   & The diffuseness of the code~smell instances in the source code of an app.\\\cline{2-3}
		& \textsf{\#removals}      & The number of instances removed in the dataset.\\\cline{2-3}
		& \textsf{\%removals}      & The percentage of instances removed---\ie $\frac{\sf \#removals}{\sf \#introductions}$.\\\cline{2-3}
		& \textsf{\#code-removed}  & The number of instances removed with source code removal.\\\cline{2-3}
		& \textsf{\%code-removed}  & The percentage of instances removed with source code removal---\ie $\frac{\sf \#code-removed}{\sf \#removals}$.\\
		\hline
		\multirow{8}{*}{\rotatebox[origin=c]{90}{\textbf{\emph{Commit}}}}
		& \textsf{\#commit-introductions} & The number of code~smell instances introduced by the commit.\\\cline{2-3}
		& \textsf{\#commit-removals}      & The number of code~smell instances removed by the commit.\\\cline{2-3}
		& \textsf{distance-to-release}    & The distance between the commit and the next release in terms of number of commits.\\\cline{2-3}
		& \textsf{time-to-release}        & The distance between the commit and the next release in terms of number of days.\\\hline
	\end{tabularx}
	\egroup
\end{table}

As shown in Table~\ref{table:code_smells}, every code~smell type affects a specific entity of the source code.
Therefore, to compute the metric \textsf{\%diffuseness}, we only focused on these entities.
For instance, the code~smell \emph{Init~OnDraw} affects only the entity \texttt{View}, thus we compute the percentage of views affected.
This allows us to focus on the relevant parts of the source code and have a precise vision about the code~smell diffuseness.
For each app $a$, the diffuseness of a type of code~smells $t$ that affects an entity $e$ is defined by: 
\[\textsf{\%diffuseness}(a,t)= \frac{\textsf{\#affected-entities}(a,t)}{\textsf{\#available-entities}(a,e)}\] 
For instance, the diffuseness of the code smell \emph{No Low Memory Resolver (NLMR)} in an app $a$ is:
\[\textsf{\%diffuseness}(a,NLMR)= \frac{\textsf{\#NLMR-instances}(a)}{\textsf{\#activities}(a)}\] 
Where $\textsf{\#NLMR-instances}(a)$ is the number of \emph{No Low Memory Resolver} instances in the app $a$ and $\textsf{\#activities}(a)$ is the number of activities in $a$.

For the metrics \textsf{\#code-removed} and \textsf{\%code-removed}, we tracked the source code modifications that led to code~smell removals.
In particular, we counted all code~smell removals where the host entity was also removed.
For example, when an instance of the code~smell \emph{No~Low~Memory~Resolver} is removed, the removal can be counted as \textsf{\#code-removed} only if the host \texttt{Activity} has also been removed in the same commit. 

\subsubsection{\textsc{RQ\,1}: How frequent and diffuse are mobile code~smell introductions?}
To inspect the prevalence of code~smells, we computed---for each code~smell type---the metrics:
\begin{inparaitem}[]
	\item \textsf{\#introductions} and
	\item \textsf{\%affected-apps}. 
\end{inparaitem}
These metrics allow us to compare the prevalence of different code~smell types.
Then, to obtain a precise assessment of this prevalence, we also used the metric:
\begin{inparaitem}[]
	\item \textsf{\%diffuseness}.
\end{inparaitem}
We computed the diffuseness of each code~smell type in every app of our dataset.
Finally, we plotted the distribution to show how diffuse are code~smells compared to their host entities.

\subsubsection{RQ\,2: How do releases impact introductions and removals of mobile code~smells?}
This research question focuses on the impact of releases on code~smell evolution.
To ensure the relevance of this investigation, we paid careful attention to the suitability of the studied apps for a release inspection.
In particular, we manually checked the timeline of each app to verify that it publishes releases through all the change history.
We excluded apps that did not use releases at all, and apps that used them only at some stage.
For instance, the \textsf{Chanu} app~\cite{chanu} only started using releases in the last $100$ commits, while the first $1,337$ commits do not have any releases.
Hence, this app is, to a large extent, release-free and thus irrelevant for this research question.
Out of the $324$ studied apps, we found $156$ that used releases during all the change history.
The list of these apps can be found in our study artifacts~\cite{CompanionArtifacts}.
It is also worth noting that as Android apps are known for continuous delivery and releasing~\cite{versioning,mcilroy2016fresh}, we considered in this analysis both minor and major releases.
This allows us to perform a fine-grained study with more releases to analyse.

We used this set of $156$ apps to evaluate the impact of releases on code~smell introductions and removals.
First, we visualized for each project the evolution of source code and code~smells along with releases.
We also plotted the evolution of code~smell diffuseness for all studied apps.
This visualization provides insights into the impact of releases and the evolution patterns of code~smells.

To accurately measure the impact of releases, we analysed the efefct of approaching releases on the numbers of introductions and removals performed in commits.
Therefore, we used the metrics \textsf{distance-to-release} and \textsf{time-to-release}.

\paragraph*{Distance to release}
We aimed to evaluate the relationship between the distance to release and the numbers of code~smells introduced and removed per commit. 
For this purpose, we assessed the correlation between the \textsf{distance-to-release} and both \textsf{\#commit-introductions} and \textsf{\#commit-removals} using Spearman's rank coefficient.
Spearman is a non-parametric measure that assesses how well the relationship between two variables can be described using a monotonic function.
This measure is adequate for our analysis as it does not require the normality of the variables and does not assess the linearity.

\paragraph*{Time to release}
Using the metric \textsf{time-to-release}, we extracted three commit sets:
\begin{compactitem}
	\item Commits authored $1$ day before a release,
	\item Commits authored $1$ week before a release,
	\item Commits authored $1$ month before a release.
\end{compactitem}

\bigskip
Then, we compared the \textsf{\#commit-introductions} and \textsf{\#commit-removals} in the three sets using Mann-Whitney U and Cliff's\,$\delta$.
We used the two-tailed Mann-Whitney U\,test~\cite{sheskin2003handbook} with a $99\,\%$ confidence level, to check if the distributions of introductions and removals are identical in the three sets.
To quantify the effect size of the presumed difference between the sets, we used Cliff's\,$\delta$~\cite{romano2006appropriate}.
Cliff is a non-parametric effect size measure, which is reported to be more robust and reliable than Cohen’s\,d~\cite{cohen1992power}.
Moreover, it is suitable for ordinal data and it makes no assumptions of a particular distribution~\cite{romano2006appropriate}. 
For interpretation, we followed the common guidelines: negligible~(N) for $|d| < 0.10$, small~(S) for $0.10 \leq |d| < 0.33$, medium~(M) for $0.33 \leq |d| < 0.474$, and large~(L) for $|d| \geq 0.474$~\cite{grissom2005effect}.

\subsubsection{RQ\,3: How do developers remove mobile code~smells?}

\paragraph{Quantitative analysis}
First, we computed for each code~smell type the metrics:
\begin{inparaitem}[]
	\item \textsf{\#removals} and
	\item \textsf{\%removals}.
\end{inparaitem}
Then, to gain insights about the actions that lead to code~smell removals, we computed:
\begin{inparaitem}[]
	\item \textsf{\#code-removed} and
	\item \textsf{\%code-removed}.
\end{inparaitem}
The metric \textsf{\%code-removed} reports the percentage of code~smell instances that were removed with source code removal.
This metric provides us a first idea about code~smell removal techniques.
To push further and identify the fine-grained actions that removed code~smells, we opted for qualitative analysis.

\paragraph{Qualitative Analysis}
The objective of our analysis is to understand how code~smells are removed. 
To achieve this, we manually analysed a sample of code~smell removals. 
We used a stratified sample to make sure to consider a statistically significant sample for each code~smell.
In particular, we randomly selected a set of $561$ code~smell removals from our dataset.
This represents a $95\,\%$ statistically significant stratified sample with a $10\,\%$ confidence interval of the $143,995$ removals detected in our dataset. 
The stratum of the sample is represented by the $8$ studied code~smells. 
This sample includes commits from 
After sampling, we analysed every smell-removing commit to inspect two aspects:
\begin{compactitem}
	\item \textbf{Commit action:} The source code modification that led to the removal of the code~smell instance.
	In this aspect, every code~smell type has different theoretical ways to remove it.
	We inspect the commits to identify the actions used in practice for concretely removing code~smells from the codebase;
	\item \textbf{Commit message:} We checked the messages looking for any mention of code~smell removal.
	In this regard, we were aware that developers could refer to the smell without explicitly mentioning its name.
	Therefore, we thoroughly read the commit messages to look for implicit mentions of the code~smell removal.
\end{compactitem}

\subsubsection{\textsc{RQ\,4}: Do developers refactor mobile code~smells?}
The objective of this question is to verify if the code~smell removals detected in the change history are actual refactoring operations.
For this purpose, we randomly selected a set of $340$ smell-removing developers---\ie developers who performed code~smell removals in our dataset.
Then, we sent them emails to ask about the removed code~smells.
In particular, we presented the concerned code~smell with the definition and code snippet that illustrates it.
Then, we asked them the following questions:
\begin{enumerate}[noitemsep]
	\item Were you aware of this code~smell?
	\item Did you refactor this code~smell intentionally?
\end{enumerate}
The objective of the first question is to capture the developer's knowledge and awareness of the code~smell.
The second question allows us to check if the code~smell removals authored by the developer are intended refactorings.
Depending on the outcome of the second question, we asked one of the following questions:
\begin{enumerate}[noitemsep]\addtocounter{enumi}{2}
	\item Why did you refactor this code~smell?
	\item Why did not you refactor this code~smell?
\end{enumerate}
These open-questions allow developers to express their thoughts about mobile code~smells and explain their choices about refactoring.

We received answers from $25$ developers, which represents a response rate of $7,35\%$.
This rate is expectedly low as we ask developers about multiple code~smell instances that impose them some deeper investment to recall and understand.
The participants answered about all studied code~smells, except \emph{Unsupported Hardware Acceleration} and \emph{Unsuited LRU Cache Size}.
None of the responding developers was involved with these two code~smells, which were indeed rare in our dataset.
While most of the respondents only answered by text, two developers showed an interest in the topic and we were able to perform online interviews with them.
The interviews initially followed the same textual questions, but depending on developers' answers, we asked additional questions.
Consequently, we were able to get more detailed answers, especially for the two open-questions.

We transcribed the interview recordings into text using a denaturalism approach, which allows us to focus on informational content while still keeping a “full and faithful transcription”.
Together, the interviews and the answers to our open questions formed material for qualitative inspection.
To analyse this material, we followed the analytical strategy of Schmidt~\emph{et al.}~\cite{schmidt2004analysis}, which is well adapted for open questions.
In this analysis, we relied on the two semantic categories:
\begin{itemize}[noitemsep]
	\item The reasons why developers refactor code~smells;
	\item The reasons why developers do not refactor code~smells;
\end{itemize}
To encode our material, we read the developers' answers and we tried to identify passages that relate to these categories.
Based on these passages, we formulated new sub-categories.
In our case, a sub-category represents a new reason for refactoring or not the code~smell.
To avoid redundancy, these sub-categories will later be presented when we report the results of this research question.

%
%
	\section{Study Results}\label{sec:results}
This section reports on the results of our study.
It is worth noting that, to facilitate the replication of this study, all the results presented here are included in our companion artifacts~\cite{CompanionArtifacts}.

\subsection{\textsc{RQ\,1}: How frequent are code~smell introductions?}
Table~\ref{table:intro_smell} reports on the number of code~smells introduced and the percentage of apps affected.

\begin{table}[!htbp]
\centering
\caption{Numbers of code smell introductions.}
\label{table:intro_smell}
\resizebox{\columnwidth}{!}{%
\begin{tabular}{l*{8}{c}c}
\hline
\emph{\textbf{Code smell}} & \emph{\textbf{LIC}} & \textit{\textbf{MIM}} & \emph{\textbf{NLMR}} & \emph{\textbf{HMU}} & \emph{\textbf{UIO}} & \emph{\textbf{UHA}} & \emph{\textbf{IOD}} & \emph{\textbf{UCS}} & \emph{\textbf{All}}\\ \hline

\textbf{\textsf{\#introductions}} & 98,751 & 72,228 & 4,198 & 3,944 & 514 & 267 & 93  & 18  & 180,013 \\ \hline
\textbf{\textsf{\%affected-apps}} &     96 &     85 &    99 &    60 &  36 &  20 & 15  &  2  &      99 \\ \hline
\end{tabular}
}
\end{table}

%

The table shows that, in the $324$ analysed apps, $180,013$ code~smell instances were introduced.
This number reflects the widespread of code~smells in Android apps.
Nonetheless, not all code~smells are frequently introduced.
Indeed, the table shows a significant disparity between the different code~smell types.
The code~smells \emph{Leaking Inner Class} and \emph{Member Ignoring Method} were introduced more than $70,000$ times, while \emph{Unsuited~LRU~Cache~Size} and \emph{Init OnDraw} were only introduced less than 100~times.
These results highlight two interesting observations:
\begin{compactitem}
	\item The most frequently introduced code~smells, \emph{Leaking Inner Class} and \emph{Member Ignoring Method}, are both about source code entities that should be static for performance optimization;
	\item The UI-related code~smells (\emph{UI Overdraw}, \emph{Unsupported~Hardware~Acceleration}, and \emph{Init OnDraw}) are among the least frequently introduced code~smells.
\end{compactitem}

Regarding affected apps, Table~\ref{table:intro_smell} shows that $99\%$ of apps had at least one code~smell introduction in their change history, which again highlights the widespread of the phenomenon.
The table also shows that the disparity in introduction frequency is reflected in the percentage of affected apps as frequent code~smells tend to affect more apps.
However, we observe that having more instances does not always imply affecting more apps.
In particular, \emph{No Low Memory Resolver} is much less present than \emph{Leaking Inner Class} and \emph{Member Ignoring Method}, $4,198$ \emph{vs.} $98,751$ and $72,228$, respectively.
Yet, it affected more apps, $99\,\%$ \emph{vs.} $96\,\%$ and $85\,\%$.

\begin{figure}[!htbp]
	\centering
	\includegraphics[width=.8\linewidth]{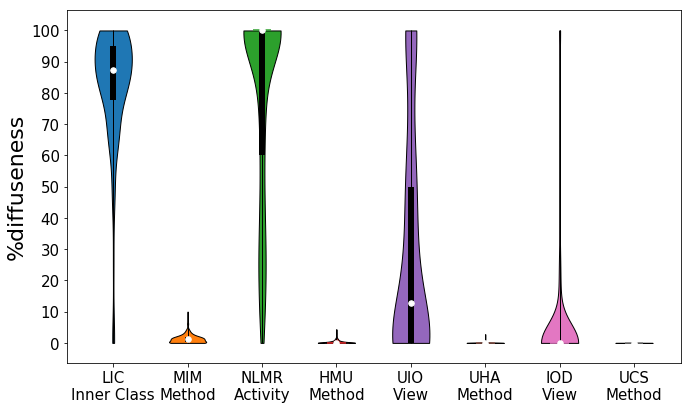}
	\caption{Distribution of code~smell \textsf{\%diffuseness} in studied apps.}
	\label{fig:diffuseness}
\end{figure}

To obtain a clear vision about these disparities, we reported in Figure~\ref{fig:diffuseness} the diffuseness of code~smells within their host entities in the studied apps.
The figure shows that \emph{No Low Memory Resolver} is the most diffuse code~smell.
At least $50\,\%$ of the dataset apps had this code~smell in all their activities, $median=100\%$.
\emph{Leaking Inner Class} is also very diffuse.
In most of the apps, it affected more than $80\,\%$ of inner classes.
Code~smells that are hosted by views are less diffuse.
On average, $15\,\%$ of the views are affected by \emph{UI Overdraw}.
As for \emph{Init OnDraw}, generally, it only affected less than $10\,\%$ of the views.
Finally, code~smells hosted by methods are the least diffuse.
\emph{Member Ignoring Method}, \emph{HashMap Usage}, \emph{Unsupported~Hardware~Acceleration}, and \emph{Unsuited~LRU~Cache~Size} are present in less than $3\,\%$ of the methods.
This low diffuseness is not surprising as the number of methods is very high.

These results show that some frequent code~smells, like \emph{Member Ignoring Method}, are not diffuse, they only impact a small proportion of their potential host entities.
Yet, code~smells that seem less frequent, like \emph{UI Overdraw} and \emph{Init OnDraw}, are more diffuse and affect a bigger proportion of entities.\\

\begin{mdframed}
	{\bf Android code~smells are not introduced and diffused equally. \emph{No~Low~Memory~Resolver} and \emph{Leaking~Inner~Class} are the most diffuse, in average they impact more than $80\,\%$ of the activities and inner classes, respectively.}
\end{mdframed}

\subsection{\textsc{RQ\,2}: How do releases impact introductions and removals of mobile code~smells?}\label{subsection:releases}
%
%

\begin{figure}
	\begin{subfigure}{0.5\textwidth}
		\centering
		\includegraphics[width=\textwidth]{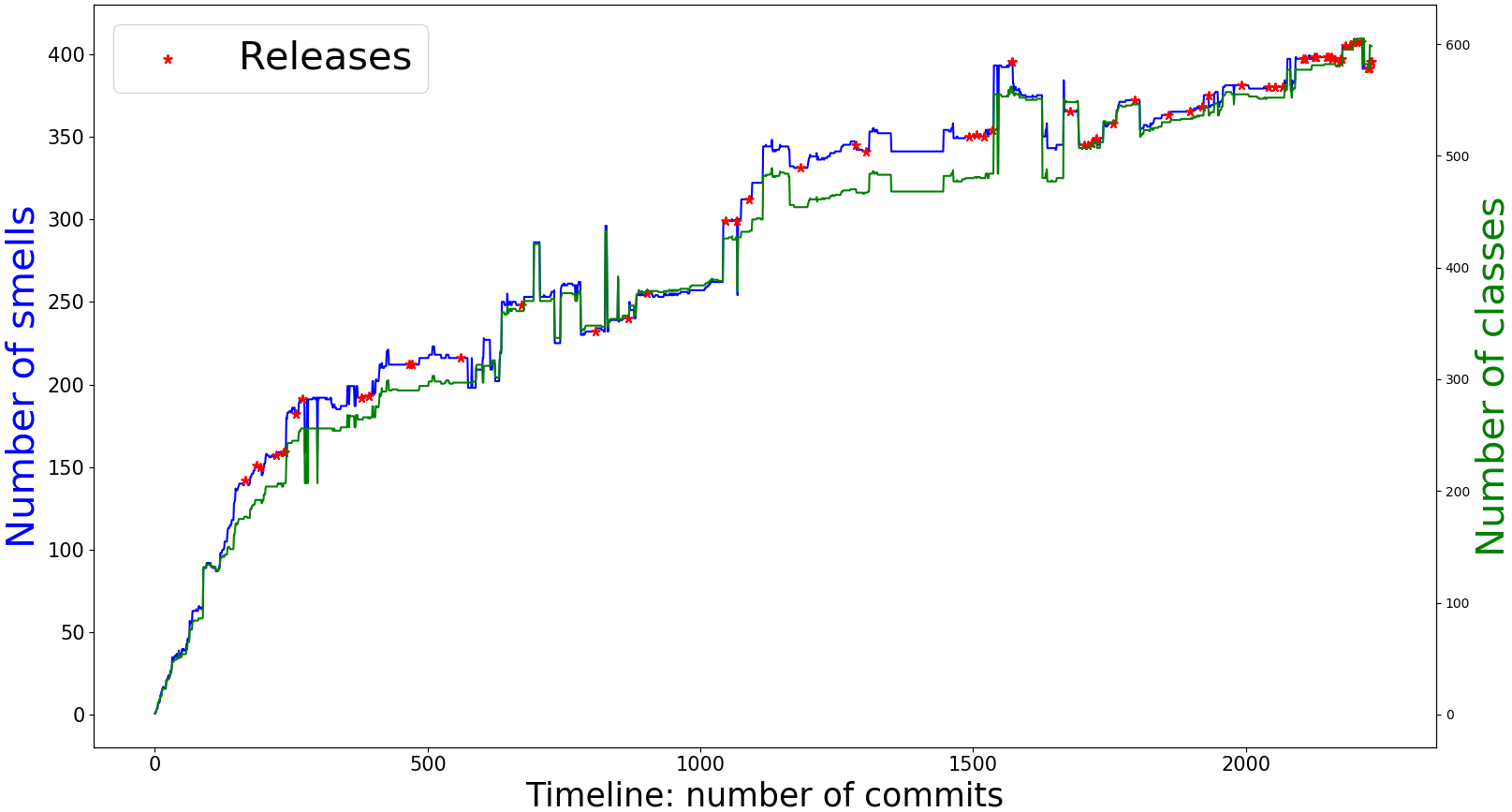}
		\caption{\textsf{Seafile}}
		\label{fig:seadroid}
	\end{subfigure}
	\begin{subfigure}{0.5\textwidth}
		\centering
		\includegraphics[width=\textwidth]{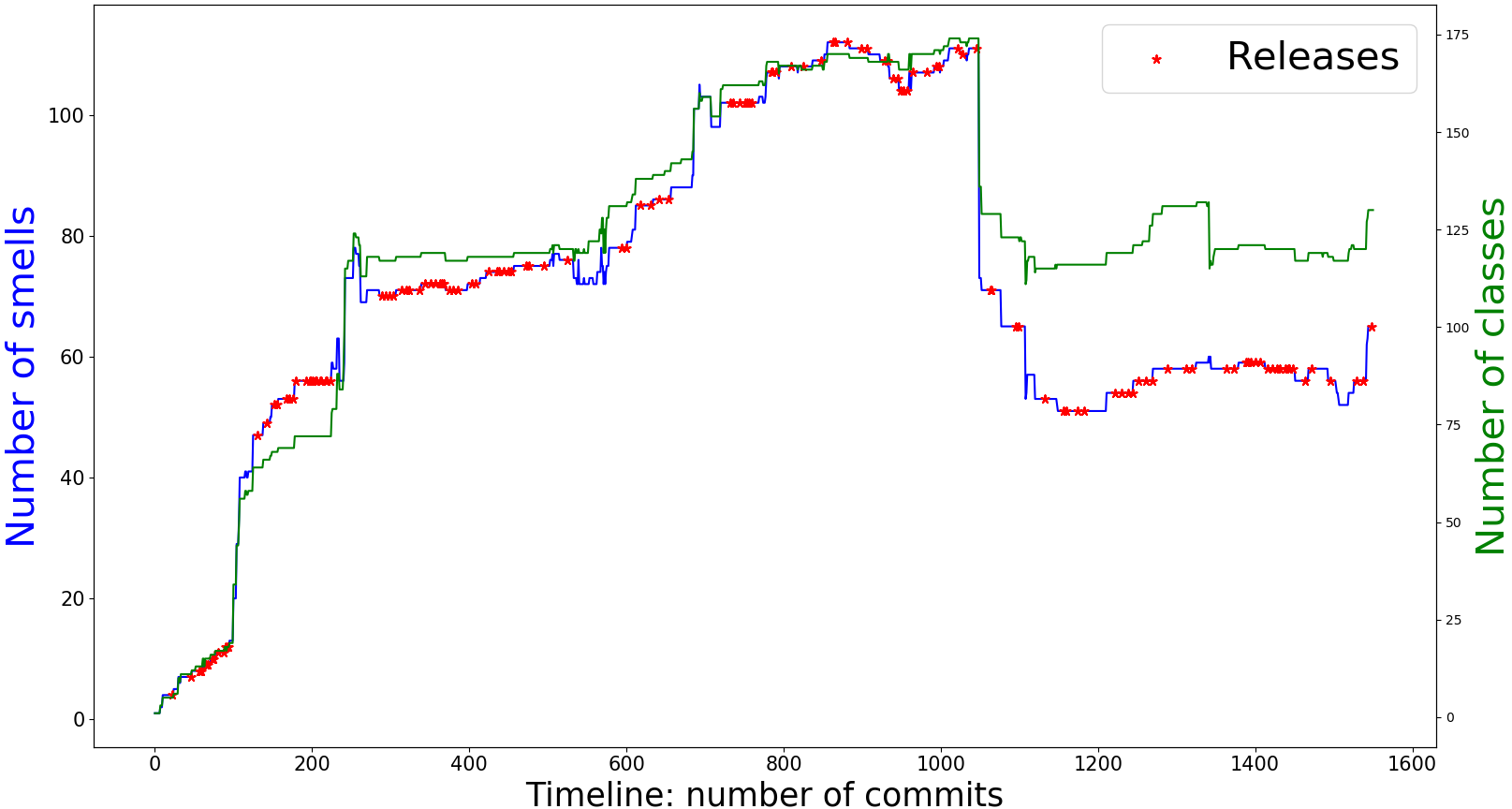}
		\caption{ \textsf{Syncthing}}
		\label{fig:syncthing}
	\end{subfigure}

	\begin{subfigure}{0.5\textwidth}
		\centering
		\includegraphics[width=\textwidth]{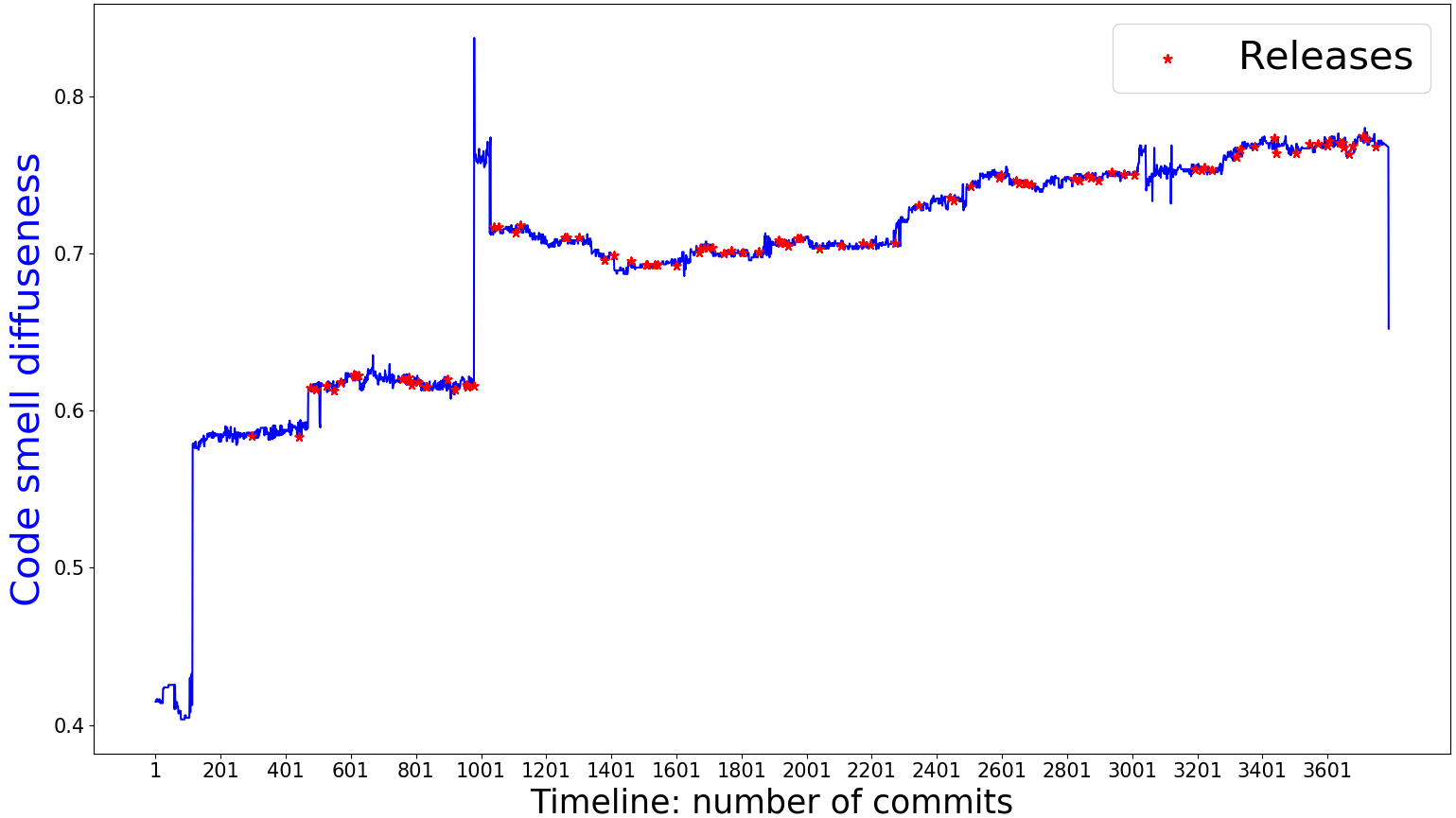}
		\caption{ \textsf{Subsonic}}
		\label{fig:Subsonic}
	\end{subfigure}
	\begin{subfigure}{0.5\textwidth}
		\centering
		\includegraphics[width=\textwidth]{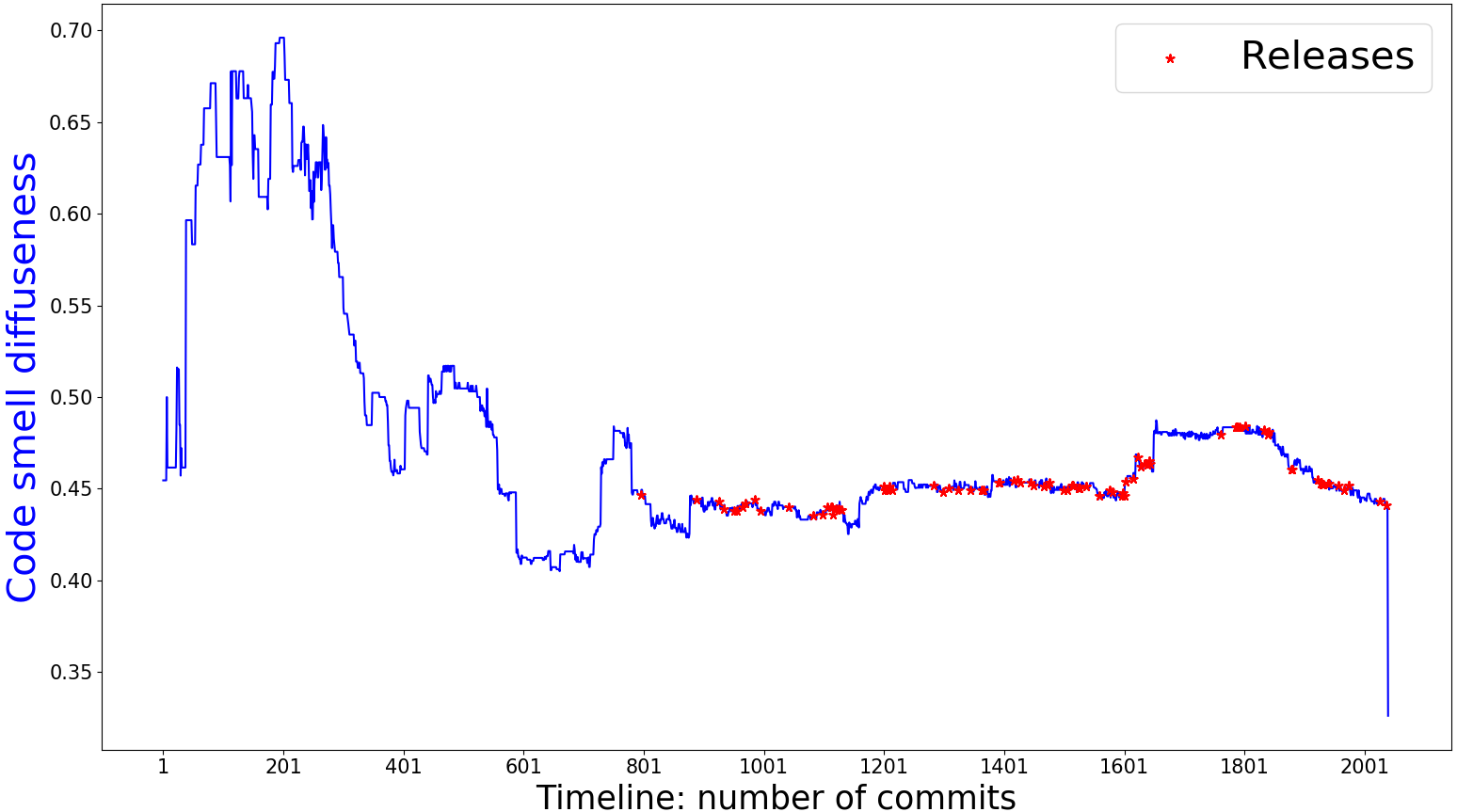}
		\caption{ \textsf{Andstatus}}
		\label{fig:andstatus}
	\end{subfigure}

	\begin{subfigure}{0.5\textwidth}
		\centering
		\includegraphics[width=\textwidth]{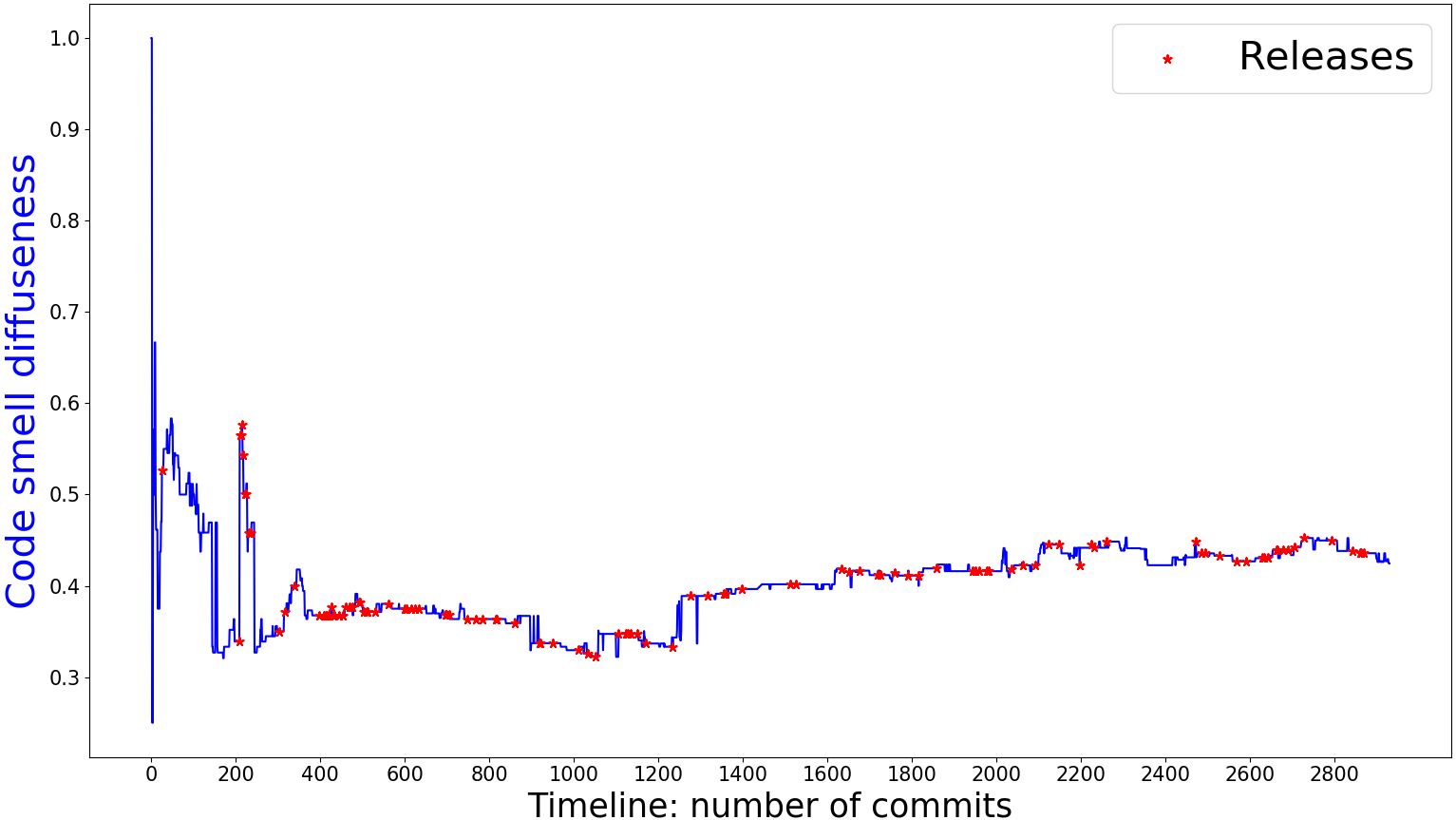}
		\caption{\textsf{KISS}}
		\label{fig:KISS}
	\end{subfigure}
	\begin{subfigure}{0.5\textwidth}
		\centering
		\includegraphics[width=\textwidth]{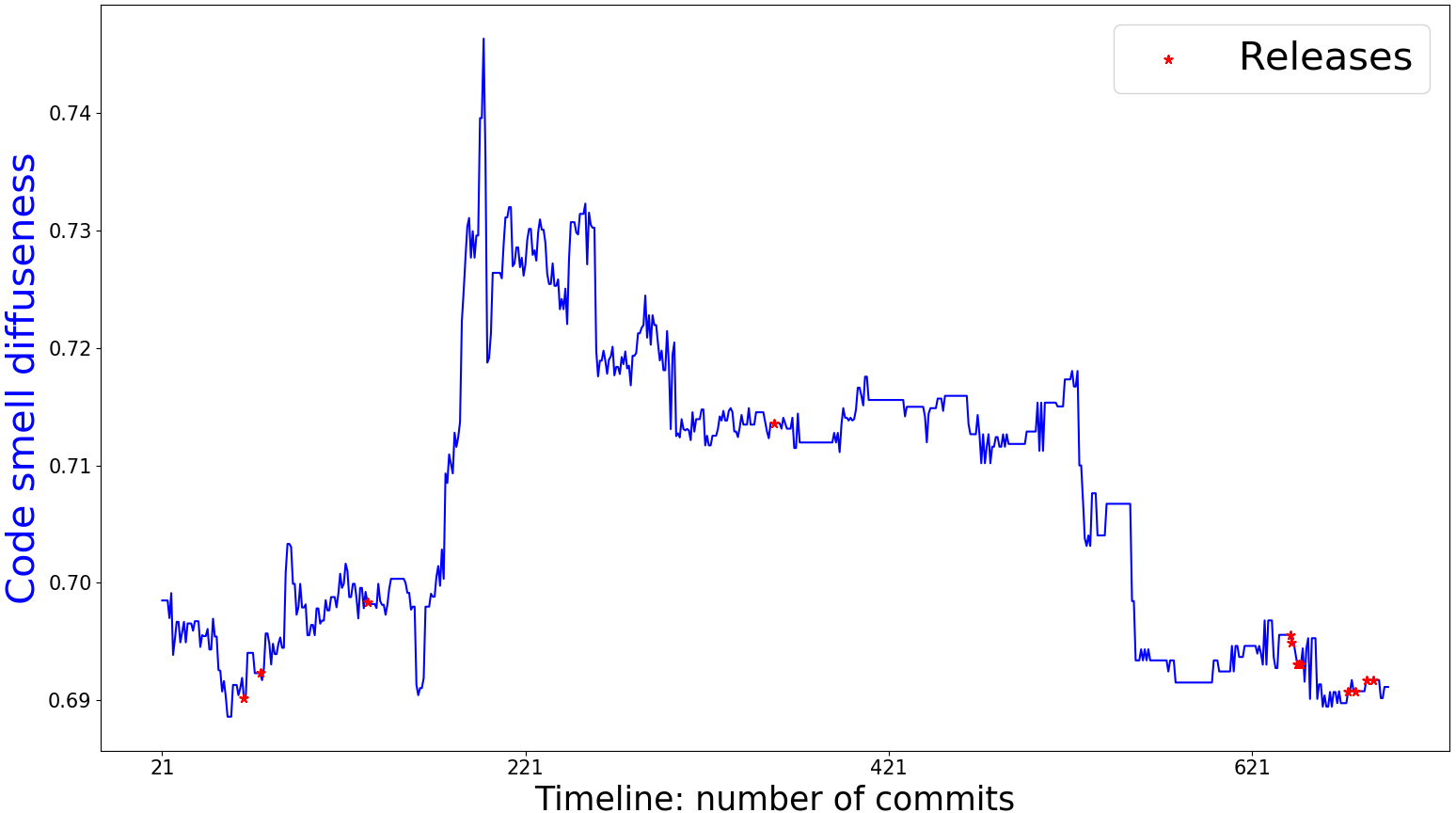}
		\caption{\textsf{4pdaClient}}
		\label{fig:4pdaClient}
	\end{subfigure}
\caption{The evolution of code smells in different Android apps.}\label{fig:evolution}
\end{figure}

In this section, we report on the results of our release analysis on the $156$ apps that used releases regularly.
For each app, we generated code~smell evolution curves that can be found in our artifacts~\cite{CompanionArtifacts}.
Figure~\ref{fig:seadroid} shows an example of these curves that depicts the evolution of the number of code~smells and classes in the \textsf{Seafile} client app.
The figure highlights the releases to show the changes in code~smell numbers when approaching releases.
From our manual examination of all the evolution curves, we did not observe any tendency of code~smell increase or decline immediately before or after releases.
Generally, the number of code~smells evolves with an important growth at the first stages of feature development.
Then, this growth stabilizes as the projects enter the maintenance phase.
Naturally, this pattern is not followed by all the analysed projects as in many cases some components or modules are removed, which results in a drop in the project size and the number of code~smells.
Figure~\ref{fig:syncthing} presents an example of these cases that were observed in the \textsf{Syncthing} app.
Regardless of the growth pattern, we observe that the number of code~smells follows the project size in terms of number of classes.
These observations align with Lehman's laws of continuing growth and declining quality where the increase in code~smells is an indicator of declining quality. 

To isolate the impact of project size, we also generated evolution curves for code~smell diffuseness.
Figures~\ref{fig:Subsonic}-\ref{fig:4pdaClient} show examples of curves generated for four Android apps.
From our inspection of these curves, we did not observe any impact of releases on code~smell diffuseness.
Sometimes, we notice abrupt drops or peaks in code~smell diffuseness but these events are not explicitly related to releases.
We also notice that the diffuseness evolution did not follow one simple pattern, like the raw number of code~smells.
However, based on general trends, we observed that three patterns were emerging frequently: consistent rise, consistent decline, and stability.

Figure~\ref{fig:Subsonic} shows an example of consistent rise in code~smell diffuseness observed in the \textsf{Subsonic} project.
We can see how the project started with $0.4$ code~smells per class and rose consistently to reach $0.8$ code~smells per class after $3600$ commits.
The opposite pattern is observed in Figure~\ref{fig:andstatus} where code~smell diffuseness declines over the lifetime of the \textsf{AndStatus} app.
At the early stages of this project, the diffuseness was around $0.65$ smells per class and it declined progressively and ended up around $0.4$ smells per class.
The \textsf{KISS} app, depicted in Figure~\ref{fig:Subsonic}, shows an example of stable code~smell diffuseness.
Despite some abrupt peaks and drops in the initial commits, code~smell diffuseness always ranged  between $0.35$ and $0.45$ all along $2800$ commits.
Some apps did not fall under any of these patterns and their code~smell evolution had random changes along the project lifetime.
For instance, the \textsf{4pdaClient} app, had a hill-shaped evolution curve as shown in Figure~ \ref{fig:4pdaClient}. 
Indeed, the diffuseness evolved constantly in the first $200$ commits, then started decreasing to go back to the same initial diffuseness.

Beyond this manual analysis, we assessed the impact of releases using the metrics \textsf{distance-to-release} and \textsf{time-to-release}.

\subsubsection{Distance to release}
\begin{figure}
	\centering
	\includegraphics[width=.9\linewidth]{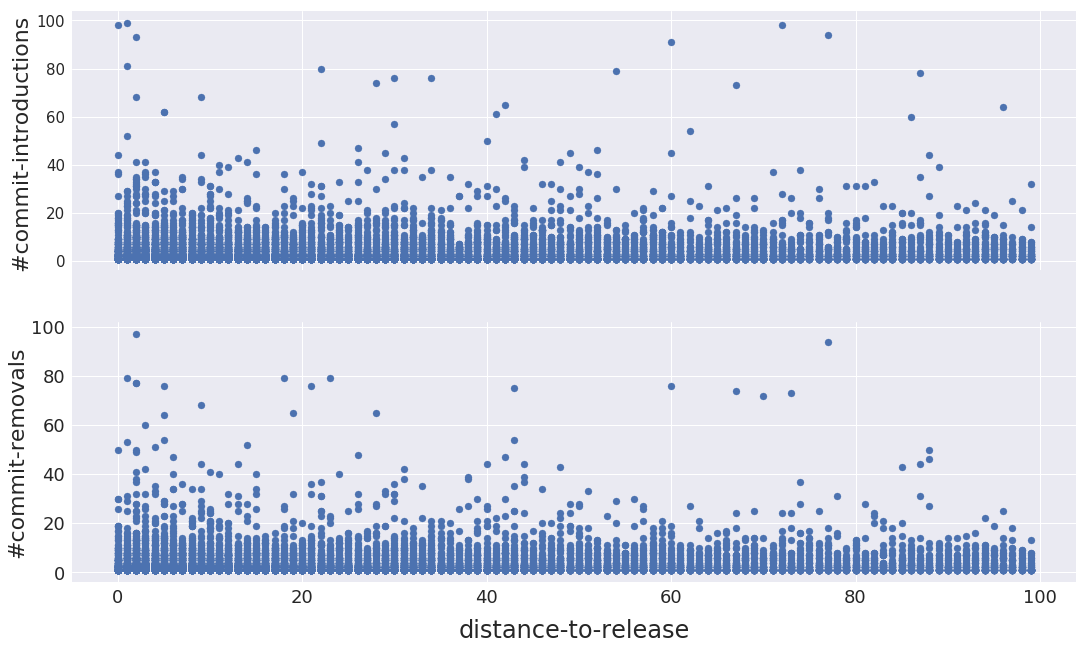}
	\caption{The number of code~smell introductions and removals per commit in the last 100 commits before release.}
	\label{fig:intros_removals_releases}
\end{figure}

Figure~\ref{fig:intros_removals_releases} presents two scatter plots that show the relationship between the distance from releasing and the number of code~smell introductions and removals per commit. 
The first thing that leaps to the eye is the similarity between the two plots.
Code~smell introductions and removals are similarly distributed regarding the distance from releasing.
We do not notice any time window where the code~smell introductions and removals are negatively correlated.
We also do not visually observe any correlation between the distance from release and code~smell introductions and removals.
Indeed, the Spearman's rank correlation coefficients confirm the absence of such correlations.
\small
$$
Spearman (\text{\sf{distance-to-release}}, \text{\sf{\#commit-introductions}}) \left\{
\begin{array}{ll}
\rho= 0.04 \\
\emph{p-value}<0.05
\end{array}
\right.\\
$$

$$
Spearman (\text{\sf{distance-to-release}}, \text{\sf{\#commit-removals}}) \left\{
\begin{array}{ll}
\rho= 0.01 \\
\emph{p-value}<0.05
\end{array}
\right.
$$

The results show that for both correlations, the \emph{p-value} is below the threshold. 
Hence, we can consider the computed coefficients as statistically significant.
As these correlation coefficients are negligible, we can conclude that there is no monotonic relationship between the distance from releasing and the numbers of introductions and removals per commit. 

\subsubsection{Time to release}
After analysing the impact of the distance to release, we investigate the impact of the time to release on code~smell introductions and removals.

\begin{figure}
	\centering
	\includegraphics[width=\linewidth]{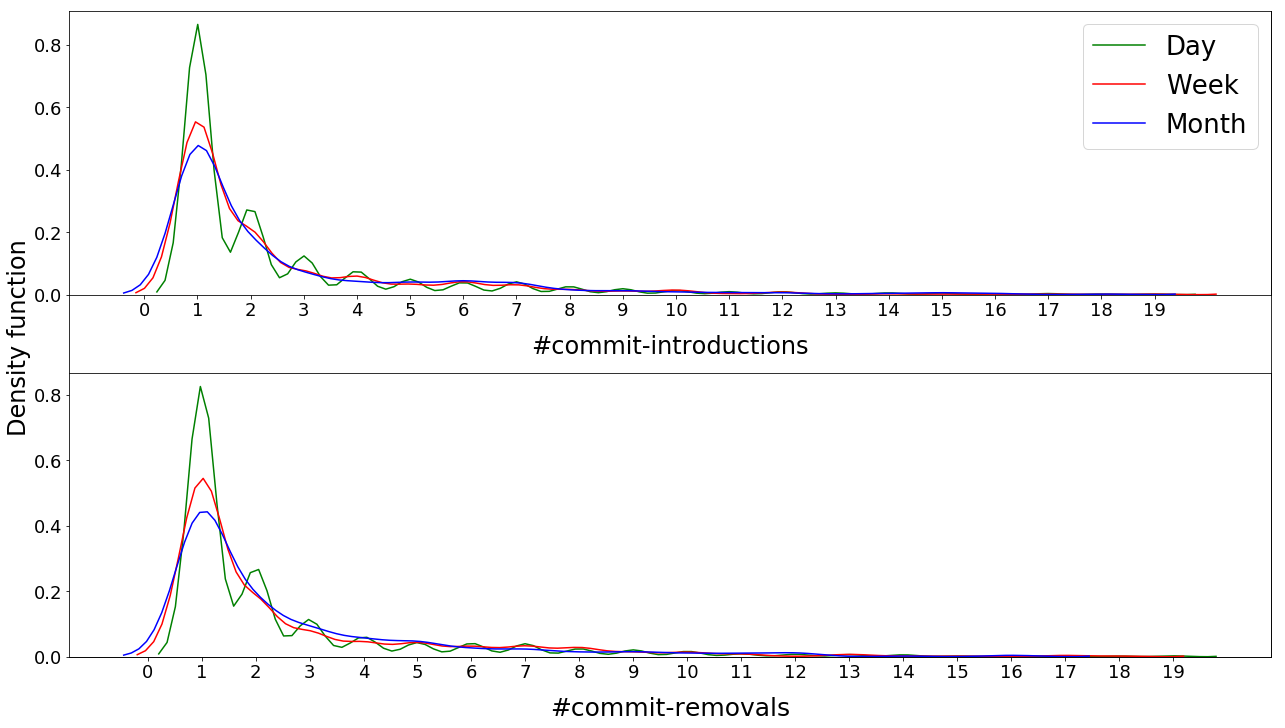}
	\caption{The density function of code~smell introductions and removals one day, one week, and one month before releasing.}
	\label{fig:density_I_R}
\end{figure}

Figure~\ref{fig:density_I_R} shows the density function of code~smell introductions and removals in different timings.
First, we observe that code~smell introductions and removals are distributed similarly.
For each timing, the density function of code~smell introductions and removals are analogous.

As for the comparison between code~smell introductions performed at different times, we observe that commits performed one day before releasing have a higher probability to only have one code~smell introduction.
Commits performed one week or one month before release tend also to have around one code~smell introduction, but they also have chances to introduce more code~smells.
This means that commits authored one day before the release do not necessarily have more code~smell introductions.  

Code~smell removals follow the same distribution for every timing.
Thus, we can infer that time to release has no visible impact on code~smell introductions and removals.

To confirm this observation, we compare in Table~\ref{table:compare_release_introductions} the code~smell introductions performed one day, one week, and one month before the release.

\begin{table}[!htbp]
	\centering
	\caption{Compare \textsf{\#commit-introductions} in commits authored one day, one week, and one month before releasing.}
	\label{table:compare_release_introductions}
	\resizebox{\columnwidth}{!}{%
		\begin{tabular}{l*{9 }{c}}
			\cline{2-10}
			\multicolumn{1}{c}{}& \emph{\textbf{LIC}} & \emph{\textbf{MIM}} &\emph{\textbf{NLMR}}  &  \emph{\textbf{HMU}} & \emph{\textbf{UIO}} & \emph{\textbf{IOD}} & \emph{\textbf{UHA}}   &  \emph{\textbf{UCS}} &  \emph{\textbf{All}} \\ \hline

			\emph{\textbf{Day}}  & $p>0.01$ & $p>0.01$ & $p>0.01$& $p>0.01$ & $p>0.01$ & $p>0.01$ & $p>0.01$ & $-$  & $p>0.01$ \\ 
			\emph{\textbf{Week}}  & $0.01 (N)$ & $0.05 (N)$ & $0.08 (N)$ & $0.20 (S)$ & $0.18 (S)$ & $0.10 (S)$ & $-$ & $-$ & $0.00 (N)$\\ \hline
			
			\emph{\textbf{Day}}  & $p>0.01$ & $p>0.01$ & $p>0.01$& $p>0.01$ & $-$ & $-$ & $-$ & $-$ & $p>0.01$ \\ 
			\emph{\textbf{Month}}  & $0.02 (N)$ & $0.04 (N)$ & $0.08 (N)$ & $0.02 (N)$ & $-$ & $-$ & $-$ & $-$ & $0.01 (N)$\\ \hline
			
			\emph{\textbf{Week}}  & $p>0.01$ & $p>0.01$ & $p>0.01$& $p>0.01$ & 
			$-$ & $-$ & $-$ & $-$ &  $p>0.01$ \\ 
			\emph{\textbf{Month}}  & $0.04 (N)$ & $0.00 (N)$ & $0.00 (N)$ & $0.04 (N)$ & $-$ & $-$ & $-$ & $-$ & $0.01 (N)$ \\ \hline
		\end{tabular}
	}
\end{table}

The table results show that for all code~smells, there is no significant difference between code~smell introductions occurring on different dates before the release ($p-value>0.01$).
The effect size values confirm the results, all the quantified differences are small or negligible.

Similarly, Table~\ref{table:compare_release_removals} compares code~smell removals in commits authored one day, one week, and one month before the release.
The results are similar to the ones observed for code~smell introductions.
The differences between different commits sets are insignificant ($\emph{p-value}>0.01$) and effect sizes are small or negligible regardless of the code~smell type.

These observations suggest that there is no difference between the introduction and removal tendencies in commits authored just before release and those written days or weeks before.
It is worth noting that for \emph{UI Overdraw}, \emph{Init OnDraw}, \emph{Unsupported~Hardware~Acceleration}, and \emph{Unsuited~LRU~Cache~Size}, the number of instances was in some cases insufficient for performing the statistical tests.
Hence, our results are not applicable to these code~smells.
\begin{table}[!htbp]
	\centering
	\caption{Compare \textsf{\#commit-removals} in commits authored one day, one week, and one month before releasing.}
	\label{table:compare_release_removals}
	\resizebox{\columnwidth}{!}{%
		\begin{tabular}{l*{9}{c}}
			\cline{2-10}
			\multicolumn{1}{c}{}& \emph{\textbf{LIC}} & \emph{\textbf{MIM}} &\emph{\textbf{NLMR}}  &  \emph{\textbf{HMU}} & \emph{\textbf{UIO}} & \emph{\textbf{IOD}} & \emph{\textbf{UHA}}   &  \emph{\textbf{UCS}} &  \emph{\textbf{All}} \\ \hline
			
			\emph{\textbf{Day}}   & $p>0.01$ & $p>0.01$ & $p>0.01$& $p>0.01$ & $p>0.01$ & $p>0.01$ & $p>0.01$ & $-$ & $p>0.01$\\ 
			\emph{\textbf{Week}}  & $0.05 (N)$ & $0.03 (N)$ & $0.02 (N)$ & $0.05 (N)$ & $0.29 (S)$ & $-$ & $-$ & $-$ & $0.01 (N)$\\ \hline
			
			\emph{\textbf{Day}}   & $p>0.01$ & $p>0.01$ & $p>0.01$& $p>0.01$ & $-$ & $-$ & $-$ & $-$ &  $p>0.01$ \\ 
			\emph{\textbf{Month}} & $0.02 (N)$ & $0.07 (N)$ & $0.30 (S)$ & $0.01 (N)$ & $-$ & $-$ & $-$ & $-$ & $0.03 (N)$\\ \hline
			
			\emph{\textbf{Week}}  & $p>0.01$ & $p>0.01$ & $p>0.01$& $p>0.01$ & $-$ & $-$ & $-$ & $-$ & $p>0.01$ \\ 
			\emph{\textbf{Month}} & $0.02 (N)$ & $0.04 (N)$ & $0.30 (S)$ & $0.04 (N)$ & $-$ & $-$ & $-$ & $-$ & $0.01 (N)$\\ \hline
			
		\end{tabular}
	}
\end{table}

\begin{mdframed}
	{\bf Releases do not have an impact on the introductions and removals of Android code~smells.}
\end{mdframed}

\subsection{\textsc{RQ\,3}: How do developers remove mobile code~smells?}
\subsubsection{Quantitative Analysis}
Table~\ref{table:remov_smell} reports on the number and percentage of removals.
The table shows that on average $79\%$ of code~smell instances are removed. 
Looking at every code~smell type separately, the removal rate varies between $35\,\%$ and $93\,\%$.
\emph{Member Ignoring Method} is the most removed code~smell with $93\,\%$ of its instances were removed along with the change history.
On the other hand, \emph{Unsupported~Hardware~Acceleration} is the least removed code~smell with only $35\,\%$ of its instances removed.
The other code~smells have a coherent removal percentage; they all had from $60\%$ to $70\%$ of their instances removed.

\begin{table}[!htbp]
	\centering
	\caption{Number and percentage of code smell removals.}
	\label{table:remov_smell}
	\resizebox{\columnwidth}{!}{%
		\begin{tabular}{l*{8}{c}c}
			\hline
			\emph{\textbf{Code smell}} & \emph{\textbf{LIC}} & \emph{\textbf{MIM}} &\emph{\textbf{NLMR}}  &  \emph{\textbf{HMU}} & \emph{\textbf{UIO}} & \emph{\textbf{IOD}} & \emph{\textbf{UHA}}   &  \emph{\textbf{UCS}} &  \emph{\textbf{All}} \\ \hline
			
			\textsf{\textbf{\#removals}}     & 70,654 & 67,777 & 2,526 & 2,509 & 305 & 147 & 66 & 11 & 143,995 \\\hline
			\textsf{\textbf{\%removals}}     &     71 &     93 &    60 &    63 &  59 &  35 & 70 & 61 &      79 \\ \hline
			\textsf{\textbf{\#code-removed}} & 67,169 & 13,809 & 1,625 & 1,824 & 273 & 123 & 33 &  8 &  84,864 \\\hline
			\textsf{\textbf{\%code-removed}} &     95 &     20 &    64 &    73 &  90 &  84 & 50 & 73 &      59 \\\hline
		\end{tabular}
	}
\end{table}

The table also reports the number and percentage of code~smell instances removed within source code removal---\ie \textsf{code-removed}.
The table shows that overall $59\,\%$ of code~smell removals are a result of removing source code.
For all code~smell types, except \emph{Member Ignoring Method}, more than $50\%$ of code~smell removals are accompanied with the removal of their host entities.
\emph{Member Ignoring Method} is the only code~smell that is rarely removed with source code removals--- \textsf{\%code-removed}=$20\,\%$.
\begin{mdframed}
{\bf Overall, $79\%$ of code~smell instances are removed through the change history.
Except for \emph{Member Ignoring Method}, most of code~smells are removed because of source code removal.}
\end{mdframed}

\subsubsection{Qualitative Analysis}
Table~\ref{table:manual} summarizes the results of our manual analysis of $561$ smell-removing commits.
For each code~smell type, the table presents the number of analysed instances, a breakdown of the actions used to remove it, and the percentage of messages mentioning its removal. 
The following subsections report on these results in details.
For some code~smells, the removal actions were similar, thus we report them together.  
Also, for the sake of clarification, we remind for each code~smell all actions that can be performed to remove it before reporting the actions found in the analysed sample.

\begin{table}[!htbp]
	\centering
	\caption{Results of manual analysis.}
	\label{table:manual}
	
	\bgroup	
	\def\arraystretch{2}
		\begin{tabular}{l*{4}{c}}
			\hline
			\emph{\textbf{Code smell}} & \emph{\textbf{\#Instances}} & \emph{\textbf{Commit actions }} & \emph{\textbf{Message}} 
			\\ \hline
			\emph{\textbf{LIC}} & 96 & \makecell{Remove inner-class ($98\%$)\\ Make inner-class static ($2\%$)} 
			& $0\%$
			\\ \hline
			\emph{\textbf{MIM}} & 96 & \makecell{Add method body ($85\%$)\\Remove Method ($15\%$)}  
			& $0\%$
			\\ \hline
			\emph{\textbf{NLMR}}  & 93 & \makecell{Remove Activity ($65\%$)\\ Transform Activity ($35\%$)}  
			& $0\%$
			\\ \hline 
			\emph{\textbf{HMU}} & 74 & \makecell{Remove Method  ($70\%$)\\Remove statements ($30\%$)} 
			& $0\%$
			\\ \hline
			\emph{\textbf{UIO}} & 74 & \makecell{Remove Method  ($88\%$)\\Remove statements ($11\%$)\\ Add  \texttt{clipRect()} ($1\%$) } 
			& $0\%$
			\\ \hline
			\emph{\textbf{IOD}} & 40 & \makecell{Remove Method  ($55\%$)\\Remove statements ($45\%$)} 
			& $15\%$
			\\ \hline
			\emph{\textbf{UHA}}   & 59 & \makecell{Remove Method  ($85\%$)\\Remove statements ($15\%$)} 
			& $0\%$
			\\ \hline
			\emph{\textbf{UCS}}  & 11 & \makecell{Remove Method  ($91\%$)\\Remove statements ($9\%$)} 
			& $0\%$
			\\ \hline

\end{tabular}

\egroup
\end{table}

\paragraph*{Code~smell: Leaking~Inner~Class (LIC)} 
\leavevmode\newline
\textbf{Possible removals:} 
\begin{itemize}[wide=10pt,noitemsep,topsep=0pt]
	\item Make the inner class static;
	\item Remove the inner class.
\end{itemize}
\textbf{Commit actions:}
We found that in $98\,\%$ of the cases, \emph{LIC} instances are removed because inner classes are removed with other parts of the code.
For instance, a commit from the \textsf{Seadroid} app that fixes bugs also removes unused code that contained a non-static inner class~\cite{removeLIC}.
Hence, the commit has removed a \emph{LIC} instance as a side effect of the bug fixing.
This finding explains the high percentage of \textsf{code-removed} found for \emph{LIC} in the quantitative analysis ($95\,\%$).\\
We only found one case of \emph{LIC} removal that was not caused by source code deletion.
It was a commit that refactored a feature and made an inner class private and static, thus removing a code~smell instance~\cite{fixLIC}.
As this commit made diverse other modifications, we could not affirm that the action was an intended code~smell refactoring.

\noindent\textbf{Commit message: }
We did not find any explicit or implicit mention of \emph{LIC} in the messages of smell-removing commits.
Moreover, the messages did not refer specifically to the removed inner classes. 
Even the unique commit that removed a \emph{LIC} instance with a modification did not mention anything about the matter in its message~\cite{fixLIC}.

\paragraph*{Code~smell: Member~Ignoring~Method (MIM)}
\leavevmode\newline
\textbf{Possible removals:} 
\begin{itemize}[wide=10pt,noitemsep,topsep=0pt]
	\item Make the affected method static;
	\item Add method body, \ie introduce code that accesses non-static attributes to the affected method;
	\item Remove the affected method.
\end{itemize}
\noindent\textbf{Commit actions:}
We found that only $15\,\%$ of \emph{MIM} removals are due to the deletion of the host methods.
In most of cases, \emph{MIM} was rather removed with the introduction of source code.
Specifically, when empty methods are developed---with instructions added inside---they do not correspond to the \emph{MIM} definition anymore, and thus code~smell instances are removed.
Also, other instances are removed from full methods with the introduction of new instructions that access non-static attributes and methods.
Finally, we did not find any case of \emph{MIM} removal that was performed by only making the method static.

\noindent\textbf{Commit message:}
We did not find any commit message that referred to the removal of \emph{MIM} instances.

\paragraph*{Code~smell: No~Low~Memory~Resolver (NLMR)}
\leavevmode\newline
\textbf{Possible removals:} 
\begin{itemize}[wide=10pt,noitemsep,topsep=0pt]
	\item Add the method \texttt{onLowMemory()} to the activity;
	\item Remove the affected activity.
\end{itemize}

\noindent\textbf{Commit actions:}
We found that $65\,\%$ of \emph{NLMR} instances are removed with source code deletion.
This deletion is caused by large modifications in the code base, like major migrations and the addition of new features.
For instance, a commit from the \textsf{Silence} app refactored the whole app to start using \texttt{Fragment} components~\cite{removeNLMR}.
One consequence of these modifications is the deletion of the \texttt{SecureSMS} activity, which used to be an instance of \emph{NLMR}.\\
As for the remaining $35\,\%$ instances, the removal was due to the conversion of host activities into other components.
For example, a commit from the \textsf{K-9} app converts an activity that was an instance of \emph{NLMR} into a fragment~\cite{convertNLMR}.
As the code~smell \emph{NLMR} is about activities, the class as fragment does not correspond to the definition anymore and thus the code~smell instance is removed.
Other than these two actions, we did not find any other ways of removing \emph{NLMR} instances.
In particular, we did not find any case where the method \texttt{onLowMemory()} is added to refactor the code~smell.

\noindent\textbf{Commit message:}
We did not find any commit message that mentioned the removal of \emph{NLMR} instances.
We found one message that mentions that the commit performs memory improvement in two classes~\cite{mentionImprovement}.
Nonetheless, these improvements were not related to the \emph{NLMR} code~smell.

\paragraph*{Code~smells: HashMap~Usage (HMU), Unsupported~Hardware~Acceleration (UHA) \& Init~OnDraw (IOD)}
\leavevmode\newline
\textbf{Possible removals:} 
\begin{itemize}[wide=10pt,noitemsep,topsep=0pt]
	\item Remove the statements that introduced them;
	\item Remove their host entities.
\end{itemize}
\noindent\textbf{Commit actions:}
In the manual analysis of these code~smells, we inspected whether the removal was due to the deletion of large code chunks or only the removal of the specific statements that caused the instance.
We found that the instances of \emph{HMU} and \emph{UHA} are usually removed with the deletion of their host methods, $70\%$ and $85\%$ respectively.
As for \emph{IOD}, there are equally instances removed with big code deletions as instances removed with only statement removals.
Looking for potential intended refactorings, we carefully examined the cases where instances are removed at a low granularity level---\ie statements. 
In all \emph{HMU} and \emph{UHA} instances, we did not find a code~smell removal that could represent an intended refactoring.
All the instances are removed as a side effect of modifications inside methods that do not specifically target the code~smell statement.
However, we found that that $22\%$ of \emph{IOD} instances are removed with precise modifications that sound like proper refactoring.
Indeed, there are $9$ \emph{IOD} instances that specifically removed the \texttt{init} statement or extracted it out of the \texttt{onDraw()} method. 
Another element that incite us to describe these modifications as intended is that they removed the linter warning suppression  of \emph{DrawAllocation}.
This shows that the developers were aware of removing a code~smell instance.

\noindent\textbf{Commit message:}
Out of the $9$ potential proper refactorings of \emph{IOD}, we found $6$ commit messages that mentioned the code~smell removal.
This confirms that the operations are intended refactorings. 
As for \emph{HMU} and \emph{UHA}, none of the analysed messages mentioned their removal.

\paragraph*{Code~smells: UI~Overdraw (UIO) \& Unsuited LRU Cache Size (UCS)}
\leavevmode\newline
\textbf{Possible removals:} 
\begin{itemize}[wide=10pt,noitemsep,topsep=0pt]
	\item Add method calls:
	\begin{itemize}[noitemsep,topsep=0pt]
		\item  \texttt{clipRect()} or \texttt{quickReject()} for \emph{UIO};
		\item  \texttt{getMemoryClass()} for \emph{UCS};
	\end{itemize}
	\item Remove the code~smell statements;
	\item Remove the host methods.
\end{itemize}
\noindent\textbf{Commit actions:}
We found that most of \emph{UIO} and  \emph{UCS} instances are removed with other parts of the source code, $88\%$ and $91\%$ respectively.
The remainder instances were removed with modifications inside methods that implied the deletion of code~smell statements.
The only exception was one \emph{UIO} instance, which was removed with the introduction of a call to the method \texttt{clipRect()}.
This modification was the only case that sounded like a proper refactoring.

\noindent\textbf{Commit message:}
None of the analysed messages mentioned the code~smells \emph{UIO} and \emph{UCS}.\\

\begin{mdframed}
	{\bf Android code~smells are usually removed with large source code removing commits that do not mention refactoring. 
	In our dataset, \emph{Init OnDraw} and \emph{UI Overdraw} are the only code~smells that were subject to apparent refactoring and only \emph{Init OnDraw} was mentioned in commit messages.}
\end{mdframed}

\subsection{\textsc{RQ\,4}: Do developers refactor mobile-specific code~smells?}
\begin{figure}
	\centering
	\includegraphics[width=\linewidth]{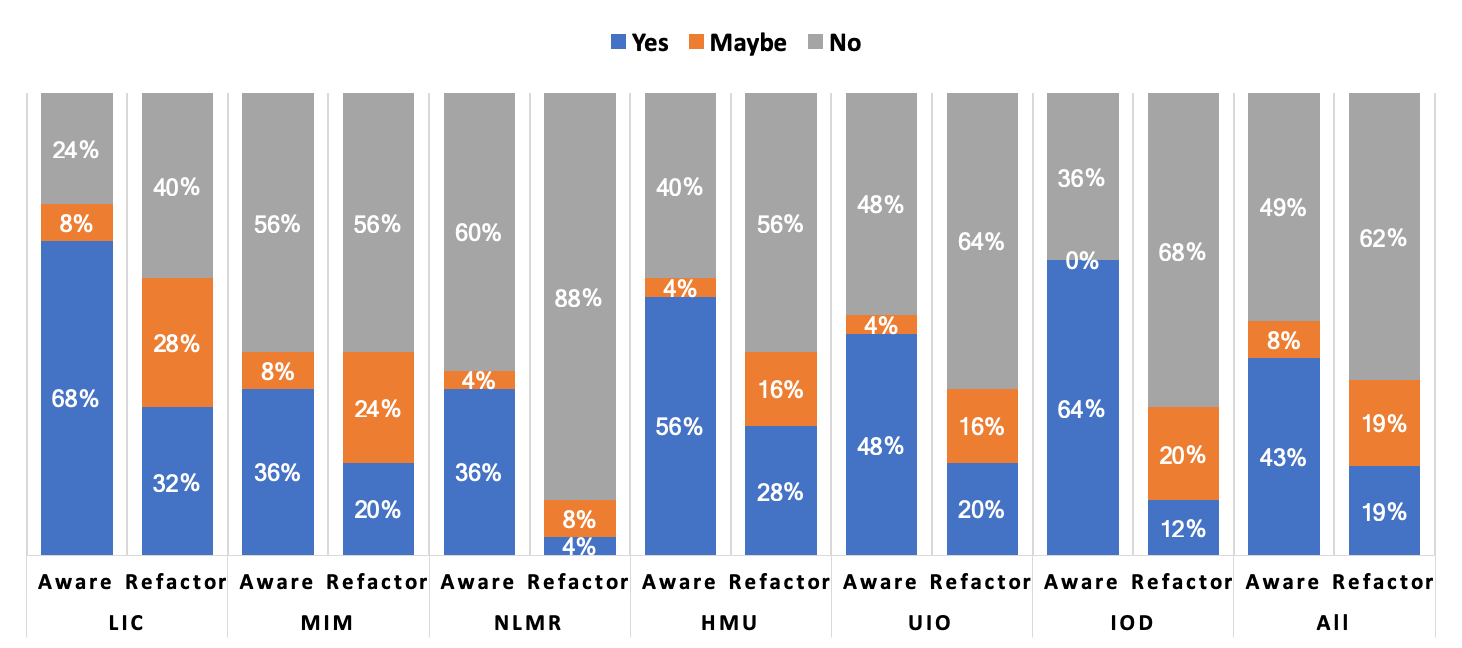}
	\caption{Answers about code~smell awareness and refactoring.}
	\label{fig:RQ4}
\end{figure}

Figure~\ref{fig:RQ4} shows the answers collected for our two first questions.
We observe that on average mobile code~smells were recognized by $43\%$ of the participants.
\emph{Leaking Inner Class}, \emph{Init OnDraw}, and \emph{HashMap Usage} are the most acknowledged code~smells.
More than $56\%$ of the participants were aware of them.
\emph{UI Overdraw} was acknowledged by $48\%$ of participants, whereas \emph{Member Ignoring Method } and \emph{No Low memory Resolver} were recognized by only $36\%$ of the participants.

When it comes to refactoring, we notice a drop by approximately $50\%$ from developers that already recognized code~smells.
Indeed, on average, only $19\%$ of the participants affirmed that they refactored mobile code~smells.
Also, \emph{Leaking Inner Class} and \emph{HashMap Usage}, which were recognized by more than $50\%$ of the participants, were only refactored by $32\%$ and $28\%$ of them.
The same drop is observed  for \emph{Member Ignoring Method} and \emph{UI Overdraw}, which were refactored by only $20\%$ of the participants.
Interestingly, \emph{No Low Memory Resolver} and \emph{Init OnDraw} were rarely refactored, $4\%$ and $12\%$ respectively.
This proportion is particularly low considering that  \emph{Init OnDraw} was acknowledged by $64\%$ of the participants.

Another observation is the large proportion of uncertain respondents for the refactoring questions---\ie developers responding with \textsf{maybe}. 
Indeed, this proportion ranged between $4\%$ and $8\%$ for the awareness question, whereas it ranged between $8\%$ and $28\%$ for the refactoring question.
This is reasonable as the participants may not remember for sure if they have already refactored this code~smell or not, thus the uncertain answer.
 
 \begin{mdframed}
 	{\bf On average, $43\%$ of the participants recognized mobile code~smells but only $19\%$ of them confirmed their refactoring.
 	Developers who are aware of Android code~smells do not necessarily refactor them. \emph{Init OnDraw} was recognized by $64\%$ of the participants and only refactored by $12\%$ of them.}
 \end{mdframed}
For the open-questions, we present the answers following the semantic sub-categories that we identified.

\subsubsection{Reasons why developers refactor mobile code~smells}

\paragraph{Code analysis tools}
Three participants claimed that they refactor code~smells because they are reported by code analysis tools.
One participant affirmed refactoring all critical code~smells that are detected by Android Lint.
\emph{``I trust the default configuration of the linter, so if something is flagged as critical, I will stop the build to fix it".}
Another participant emphasized the impact of using such tools, \emph{``IDEs and their built-in code analysis tools provide good warnings on these problems, which encourage people to fix them, even if they are unaware of them"}.
According to another participant, the help given by these tools goes beyond refactoring code~smells, \emph{``the tools provide context and explanation that sharpen the programmers' perception in the future to avoid such problems beforehand. This kind of nudging should not be underestimated"}.

 \paragraph{Personal practices}
Two developers considered code~smell refactoring as a good development practice that they adopt.
The first participant said that she refactors code~smells regularly, \emph{``I always try to improve the code source of projects that I work on. If I notice an issue, I fix it"}.
The second participant described refactoring code~smells as a part of the routine that she follows while getting into an existing code base, \emph{``for me it is a practice to dig into foreign code and while doing so, fixing smells along the way, as I gain understanding of the code base"}.
This developer explained that these smells are easy to refactor as they do not change from an app to another, \emph{``they do not relate to the app specifically, they are common knowledge"}.
 
\paragraph{Freedom}
One participant affirmed that she was able to perform refactoring operations only because she was the main maintainer of an open-source project with no external stakeholders.
\emph{``I have full control over releases. I do not have external pressure, so I can invest time in keeping the source code at the quality level that satisfies me"}.
Interestingly, this developer explicitly claims that this same freedom was not available in industrial projects that she contributed to as part of her job.

 \begin{mdframed}
	{\bf Developers who refactored Android code~smells are motivated and assisted by built-in code analysis tools and their personal commitment to code quality.}
\end{mdframed}

\subsubsection{Reasons why developers do not refactor mobile code~smells}

\paragraph{The impact is not significant}
Five participants judged that the impact of mobile code~smells is not big enough to care about them.
In particular, one developer claimed that \emph{``the performance difference is really tiny or non-existent in the end"}.
The same developer also claimed that most of these code~smells are automatically mitigated by the runtime, \emph{``ART can perform this kind of optimizations".}
Another developer explained how the architecture of mobile apps makes these code~smells less concerning, \emph{``well designed apps have most of the logic implemented in the backend and unless it's a game, the UI is not updated very often. Therefore, most of these performance issues are usually no issue at all"}.
A similar argument was constructed by another developer who believed that performance issues arise from the connection to the backend instead of the frontend code~smells, \emph{``network latency typically dominates in mobile app responsiveness"}.
Other participants gave specific examples about code~smells that seemed impotent for them.
One developer downplayed the impact of \emph{HashMap Usage}, claiming that \emph{``using SparseArray collections is better for memory usage but it is less important on modern devices than it was ten years ago"}.
\emph{No Low Memory Resolver} was also considered irrelevant by one participant who described different approaches to free memory using the activity life-cycle callbacks, \eg \texttt{onPause()} and \texttt{onResume()}.
This participant also added that this code~smell is not commonly acknowledged by developers, \emph{``I have been part of the Android community since the very beginning, 2010, and I do not recall hearing about this code~smell at all"}.

\paragraph{Not a performance problem}
Three participants expressed their doubts about the relationship between mobile code~smells and performance.
For instance, one developer considered that \emph{Member Ignoring Method} is a ``code usability" issue but not a performance one.
Another developer estimated that these code~smells are \emph{``not directly related to performance"} giving as example \emph{No Low Memory Resolver} and \emph{Leaking Inner Class}.
Another developer believed that these issues should be qualified as \emph{``code quality, completeness, correctness, resource usage issues"} instead of performance.

\paragraph{Refactoring would not help}
Two developers judged that the refactoring of code smells is useless by giving as example \emph{No Low Memory Resolver}.
Specifically, one developer considered that receiving \texttt{No Low Memory} warnings is a sign of bad memory management by the app, and responding to the system warning would not fix the issue.
She explained: \emph{``if that point is reached it probably means you have a memory leak or another problem with the way your app manages memory and clearing some cache to prevent an out of memory crash is just a temporary band-aid before the app eventually crashes anyway"}.
Another participant went further by considering that refactoring \emph{No Low Memory Resolver} could even lead to the introduction of new bugs.
She believes that \emph{``handling low memory seem more likely to do harm than good as implementations are likely to be buggy and of little benefit compared to letting the app be killed"}.

\paragraph{Performance issues are better handled reactively}
Two participants considered that developers should not worry about these code~smells because performance issues should not be handled proactively.
The first participant stated: \emph{``gut feelings about performance issues are usually wrong"}.
Thus, she advises against trusting these feelings or instincts, \emph{``never try to optimize before a profiling has shown where exactly the problem in your application is"}.
The other participant gave a similar advice and recommended relying on reactive tools like profilers to deal with performance issues when they arise, \emph{``instead of worrying about details I advise Android developers to worry more about UX and if the app performance slows just use the profiler to locate the bottlenecks and fix that"}.

\paragraph{Prioritization}
Two participants mentioned that refactoring mobile code~smells is not a high priority.
One developer referenced the perpetual trade-off between quality improvement and new features, \emph{``I always have a panoply of ideas for improving my codebase but I learned to prioritize features that directly benefit the client"}.
The developer also described different criteria that she considers while initiating a refactoring, \emph{``I evaluate considering the source code quality in the long term. If the refactoring does not help in terms of maintenance and performance, I will not perform it"}. 

\paragraph{The practice is justifiable}
One participant judged that some practices that were labeled as code~smells were justifiable.
She claimed that the use of \texttt{HashMaps} is a good and necessary practice because the alternative data structure makes code maintenance worse.
She explained this by stating that  \emph{``using Android framework \texttt{SparseArray} classes in a component prevents it from being tested in JVM unit tests"}.

\begin{mdframed}
	{\bf Developers that did not refactor Android code~smells doubt their performance impact and the usefulness of their refactoring.
	Some developers also prefer to handle performance issues when they arise instead of anticipating them.}
\end{mdframed}

\section{Discussion and Implications}\label{sec:discussion}
\paragraph{Releases}
The results of \textsc{RQ\,2} show that the pressure of releases do not have an impact on code~smell introductions and removals.
This suggests that code~smells are not introduced as a result of releasing pressure.
Moreover, when asked about the reasons for not refactoring code~smells, developers did not blame releases.
One developer mentioned prioritization, but this did not include releases and rather explained how different quality aspects and outcomes are prioritized.
These results may challenge the common beliefs about releases and their relationship with technical debt in general~\cite{tom2013exploration}.
However, it is noteworthy that our results are based on open-source projects, which can be different from their industrial counterparts.
This point was raised by the participant who praised the freedom and control that she had in her open-source project and who was aware that such circumstances are rare in industrial projects. 
Hence, we encourage future studies to:
\begin{itemize}
	\item Evaluate the impact of releases on mobile code~smells in industrial projects and ecosystems.
\end{itemize}

\paragraph{Awareness of code~smells}
Previous studies suggested that the accrual of mobile code~smells and the indifference of developers toward it are signs of unawareness~\cite{habchi2019rise,habchi2019survival}.
However, the inputs collected from developers in \textsc{RQ\,4} challenge this hypothesis.
Our participants claimed to recognize many Android code~smells, but they had other reasons to neglect them.
In particular, some developers were reluctant to code~smell refactoring because they assumed that it would lead to further issues.
To remove this obstacle, we encourage researchers and toolmakers to:
\begin{itemize}
	\item Build tools that propose automated refactoring of mobile code~smells.
\end{itemize}

\paragraph{Static analysis tools}
We observed that code~smells that are detected by Android Lint---\ie \emph{Leaking Inner Class}, \emph{Init OnDraw}, \emph{HashMap Usage}, and \emph{UI Overdraw}---are the most recognized by developers~\cite{LintCheck}.
Developers who performed refactoring also explained that their actions were motivated and assisted by built-in code analysis tools.
Also, the only apparent refactorings identified in \textsc{RQ\,3} was for \emph{Init OnDraw} and \emph{UI Overdraw}, which are detected by Android Lint.
Some of these refactorings explicitly deleted Android Lint suppressions, which shows that developers considered the linter warnings and responded with an intended refactoring.
Theses findings confirm that static analysers can help in raising awareness about code~smells and refactoring them.
Hence, we encourage researchers and toolmakers to:
\begin{itemize}
	\item Build and integrate static analysers with more code~smell coverage.
\end{itemize}

\paragraph{Removal and refactoring}
The quantitative and qualitative findings of \textsc{RQ\,3} show that code~smells are mainly removed with source code deletion.
Even for \emph{Member Ignoring Method} instances, which are removed with source code introduction, we found that they are removed because the empty and primitive methods are developed with new statements. 
While we cannot judge the intentions of a source code modification, most of the analysed commits did not reveal signs of intended refactoring and did not mention the code~smell.
On top of that, the answers collected in \textsc{RQ\,4} indicate that only a minority of smell-removing developers did perform a refactoring.
Hence, we can suggest that most of code~smell removals are a side effect of other maintenance activities and are not intentional refactorings.
This implies that:
\begin{itemize}
	\item We cannot rely on code~smell removals to learn refactoring techniques.
	Future studies that intend to learn from the change history to build automated refactoring tools cannot rely on the removals of these code~smells as learning examples.
\end{itemize}

\paragraph{Controversial  code~smells}
According to \textsc{RQ\,4}, \emph{No Low Memory Resolver} is the least acknowledged and refactored code~smell, $36\%$ and $4\%$ respectively. 
This code~smell was disapproved by many developers who explained that the absence of a resolver does not systematically result in memory issues and its presence is not always useful.
This disapproval can explain why this code~smell affected $99\%$ of the studied apps and was the most diffuse of our $8$ code~smells.
Questions were also raised about other code~smells, like UI code~smells, which were downplayed, and \emph{HashMap Usage}, which was described as justifiable and irrelevant in modern devices.
Following these questions, we invite future research works to:
\begin{itemize}
	\item Reassess the relevance of these code~smells and check the accuracy of their definitions.
\end{itemize}

\paragraph{Manage performance reactively}
Developers explained that instead of worrying about code~smells, they prefer handling performance bottlenecks when they arise.
This reactive approach was already observed and discussed in previous studies about mobile apps~\cite{habchi2018adopting,linares2015developers}, yet the research contributions in this area remain rare.
Specifically, many static analysers were provided to detect performance issues in mobile apps~\cite{habchi2017code,hecht2015detecting,palomba2017lightweight} and, to the best of our knowledge, no profiler was provided to help in managing bottlenecks when they appear.
For this reason, we encourage future works to:
\begin{itemize}
	\item Build profilers that can help developers in spotting performance bottlenecks and identifying their root causes.
\end{itemize}

\paragraph{Relationship between code~smells and performance bottlenecks}
Previous studies showed that mobile developers look after performance and take bottlenecks seriously~\cite{linares2015developers}, yet when asked about code~smells developers seem less preoccupied.
In particular, our participants doubted the impact of code~smells and questioned their association with performance.
This shows that some developers do not perceive a causal relationship between code~smells and performance bottlenecks.
Indeed, the existence of such a relationship remains theoretical.
Previous studies relied on repeated execution scenarios to demonstrate the impact of Android code~smells on performance~\cite{carette2017investigating,hecht2016empirical,palomba2019impact}, but they did not associate them with bottlenecks.
Therefore, we encourage future studies to:
 \begin{itemize}
	\item Study the relationship between mobile code~smells and performance bottlenecks.
 \end{itemize}
	
	\section{Threats to Validity}\label{sec:threats}
\paragraph{General threats}
The main threat to our internal validity could be an imprecise detection of code~smell introductions and removals. 
This imprecision is relevant in situations where code~smells are introduced and removed gradually, or when the change history is not accurately tracked.
However, this study only considered objective code~smells that can be introduced or removed in a single commit.
As for history tracking, we relied on \tandoori{}, which tracks branches and renamings and accurately detects code~smell introductions and removals ($\text{F1-score}=\{0.97,0.96\}$).
As for external validity, the main threat is the representativeness of our results.
We used a dataset of $324$ open-source Android apps with $255k$ commits and $180k$ code~smell instances. 
It would have been preferable to consider also closed-source apps to build a more diverse dataset.
However, we did not have access to any proprietary software that can serve this study. 
We encourage future studies to consider other datasets of open-source apps to extend this study~\cite{geiger2018graph,krutz2015dataset}.
We also encourage the inclusion of apps of different sizes as the frequency of mobile code smells follows the codebase size.
Another possible threat to external validity is that our study only concerns $8$ Android-specific code~smells.
Without further investigation, these results should not be generalised to other code~smells or development frameworks. 
We, therefore, encourage future studies to replicate our work on other datasets and with different code~smells and mobile platforms.

\paragraph{\textsc{RQ\,2}}
A possible threat to internal validity is the selection of releases and projects.
We avoided this threat by selecting apps that had releases all along with their change history.
This measure ensures the accuracy of the metrics \textsf{distance-to-release} and \textsf{time-to-release}.
Furthermore, our results about the impact of releases on code~smells are limited to open-source apps.
Apps developed as part of industrial projects can be subject to more external requirements and releasing pressure.
We encourage future works to extend our work by inspecting the impact of releases in different settings.

\paragraph{\textsc{RQ\,3}}
One possible threat to the internal validity of our results could be the accuracy of our manual analysis.
We tried to alleviate this threat by relying on objective criteria like the actions performed by the commit and the content of its message.
We also did not judge the intentions of developers and counted on their answers in \textsc{RQ\,4} to assess the proportion of real refactoring. 
Another threat could be the generalisability of the results of our qualitative analysis.
We used a randomly selected set of $561$ smell-removing commits.
This represents a $95\,\%$ statistically significant stratified sample with a $10\,\%$ confidence interval of the $143,995$ removals detected in our dataset. 
To support the credibility of our study, we also provide this set with our study artifacts.

\paragraph{\textsc{RQ\,4}}
The results of our user study can be threatened by the sampling bias.
We sent our questions to a set of $340$ smell-removing developers because our objective was to check if their removals were actual refactoring operations.
Without further investigation, the observed proportions of awareness and refactoring cannot be generalised to all mobile developers. 
Furthermore, the answers collected in this study may be subject to acquiescence and desirability biases.
Participants may be inclined to answer with ``yes" to agree with us or seem more aware of software quality issues.
We minimized these biases by keeping the answers anonymous and avoiding the implication that some answers are ``right" or ``wrong".
We also allowed participants to express their points of view through the open-questions.

	\section{Related Works}\label{sec:related_work} 
In this section, we report on the literature related to code~smells in mobile apps and their analysis in the change history.

\subsection{Mobile Code~Smells}
The first reference to mobile-specific code~smells was when Reimann~\emph{et~al.}~\cite{Reimann2014} proposed a catalogue of $30$ quality smells dedicated to Android.
These code~smells originate from the good and bad practices presented online in Android documentation.
They cover various aspects like implementations, user interfaces, or database usages and they are reported to harm properties, such as efficiency, user experience, or security.
Many research works built on this catalogue and proposed approaches and tools for detecting code~smells in mobile apps~\cite{habchi2017code,hecht2015detecting,kessentini2017detecting,palomba2017lightweight}.
In particular, Hecht~\emph{et~al.}~\cite{hecht2015tracking} proposed \textsc{Paprika}, a tooled approach that detects OO and Android smells in Android apps.
\textsc{Paprika} models Android apps as a large architectural graph and queries it to detect code~smells.
Palomba~\emph{et~al.}~\cite{palomba2017lightweight} proposed another tool, called \textsc{aDoctor}, able to identify $15$ Android-specific code~smells from the catalogue of Reimann~\emph{et~al.}
Habchi~\emph{et~al.}~\cite{habchi2017code} proposed an extension of \textsc{Paprika} that detects iOS-specific code~smells.
Lately, Gupta~\emph{et~al.}~\cite{gupta2019android} used $3$ machine learning algorithms to generate rules that detect four Android code~smells.
In their experiments, the \textsc{JRip} algorithm achieved the best results by generating rules capable of detecting smells with a $90\%$ overall precision.

To cope with mobile code~smells, researchers also proposed refactoring solutions.
Yu~\emph{et~al.}~\cite{lin2015refactorings} proposed \textsc{Asynchronizer}, a tool that extracts long-running operations into AsyncTasks, and \textsc{AsyncDroid}, a tool that transforms improperly-used \texttt{AsyncTasks} into Android \texttt{IntentService}.
Morales~\emph{et~al.}~\cite{morales2017earmo} proposed EARMO, an energy-aware refactoring approach for mobile apps.
They identified the energy cost of $8$ OO and mobile antipatterns.
Based on the cost, EARMO generates refactoring sequences automatically.

\subsection{Empirical Studies on Mobile Code~Smells}
Most empirical studies focused on assessing the performance impact of mobile code~smells on app performance~\cite{carette2017investigating,hecht2016empirical,morales2016anti,palomba2019impact}.
Hecht~\emph{et~al.}~\cite{hecht2016empirical} conducted an empirical study about the individual and combined impact of $3$ Android smells.
They measured the performance of $2$ apps with and without smells using the following metrics: frame time, number of delayed frames, memory usage, and number of garbage collection calls.
The measurements showed that refactoring the \emph{Member Ignoring Method} smell improves the frames metrics by $12.4\,\%$.
Carette~\emph{et~al.}~\cite{carette2017investigating} studied the same code~smells, but focused on the energy impact.
They analysed $5$ open-source Android apps and observed that in one of them the refactoring of the $3$ code~smells reduced the global energy consumption by $4,83\%$.
The study of Morales~\emph{et~al.}~\cite{morales2017earmo} also showed by analysing $20$ open-source apps that refactoring antipatterns can decrease significantly energy consumption of mobile apps.
Notably, Palomba~\emph{et~al.}~\cite{palomba2019impact} showed that methods that represent a co-occurrence of \emph{Internal Setter}, \emph{Leaking Thread}, \emph{Member Ignoring Method}, and \emph{Slow Loop}, consume $87$ times more energy than other smelly methods.

Beyond the performance impact, empirical studies compared the distribution of code~smells in mobile apps and desktop systems.
Specifically,  
Mannan~\emph{et~al.}~\cite{mannan2016understanding} compared the presence of OO code~smells in Android apps and desktop applications.
They did not observe major differences between these two types of applications in terms of density of code~smells.
However, they found that the distribution of OO code~smells in Android is more diversified than for desktop applications. 
Further, Habchi~\emph{et~al.}~\cite{habchi2017code} analysed $279$ iOS apps and $1,500$ Android apps to compare the presence of OO and mobile-specific smells in the two platforms.
They observed semantic similarities between the code~smells exhibited by the two platforms.
On top of that, they found that Android apps tend to have more OO and mobile-specific code~smells.

While these studies helped in understanding the distribution and impact of mobile-specific code~smells, they did not provide any qualitative insights about the topic.
Indeed, the only study that leveraged qualitative analysis is the one from Habchi~\emph{et~al.}~\cite{habchi2018adopting}, which investigated the perception of performance bad practices by Android developers.
This study reported that developers may lack interest and awareness about Android code~smells.
Moreover, the study showed that some developers challenge the relevance and impact of code~smells in practice.
Our work complements this study as it relies on another information source---removal instances---to understand the phenomenon of code~smells in practice.
Besides, our work also provides quantitative insights into how these code~smells are introduced and removed in practice.

\subsection{Code~smells in the Change History}
The evolution of code~smells through the change history has been addressed by various studies in the OO context.
Tufano~\emph{et~al.}~\cite{Tufano2017-TSE} addressed questions similar to our study.
They, analysed the change history of $200$ open-source projects to understand when and why code~smells are introduced and for how long they survive.
They observed that most of code~smells instances are introduced when files are created and not due to evolution process.
They also found that new features and enhancement activities are responsible for most smell introductions, and newcomers are not necessarily more prone to introducing new smells.
Interestingly, this study also investigated the rationales of code smell removal, showing that only $9\,\%$ of code~smells are removed with specific refactoring operations.
Our study yields similar results as it shows that even though
$79\,\%$ of code smell instances are removed through the change history, only $19\,\%$ of code smell removers described their actions as intentional refactoring.

Peters~\emph{et~al.}~\cite{peters2012evaluating} conducted a case study on $7$ open-source systems to investigate the lifespan of code~smells and the refactoring behaviour of developers.
They found that, on average, code~smell instances have a lifespan of approximately $50\,\%$ of the examined revisions.
Moreover, they noticed that, usually, one or two developers refactor more than the others, however, the difference is not large.
Finally, they observed that the main refactoring rationales are cleaning up dead or obsolete code, dedicated refactoring, and maintenance activities.

Tufano \emph{et al.}~\cite{tufano2016empirical} analysed the change history of $152$ open source projects to inspect the evolution of test smells and their relationship with code smells.
Their results showed that, similarly to OO code smells, test smells are introduced when tests are created and they have a high survivability.
Their results also suggest the existence of relationship between test smells and code smells of the code under test. 

In the context of mobile apps, we have already leveraged the change history to study code~smells in previous works~\cite{habchi2019rise,habchi2019survival}.
The first work studied developer contributions showing that the ownership of code~smells is spread across developers regardless of their seniority and experience.
As for the second one, it studied code~smell survival and showed that while in terms of time Android code~smells can remain in the codebase for years before being removed, it only takes $34$ effective commits to remove $75\,\%$ of them.
These results suggested that developers lack interest in code~smells and most of their actions toward them are accidental.
Result-wise, our study complements these findings as it shows the reasons behind developers' inaction toward code~smells.
Novelty-wise, our study relies on the artifacts of these works to address new topics:
\begin{compactitem}
	\item Removal fashions: This study inspects the actions that lead to code~smell removals;
	\item Refactoring: In this study, we discuss with developers to (i) check if they intentionally refactor code~smells and to (ii) identify the motivations behind their actions;
	\item Releases and diffuseness: Our previous work evaluated the impact of releases on code~smell survival~\cite{habchi2019survival}.
	In this work, we go further and assess the impact of releases on code~smell introductions and removals.
	On top of that, we provide insights about the evolution of code~smells and their diffuseness. 
\end{compactitem}

Another relevant study for our work was conducted by Mazuera-Rozo~\emph{et al.}~\cite{mazuera2020investigating} who manually analysed $500$ commits that fixed performance bugs in Android and iOS apps.
This analysis allowed them to build a taxonomy of performance bugs and confirm that GUI lagging, energy leak, and memory bloat are the most common performance bugs in mobile apps.
The study also analysed the survival of performance bugs showing that on average they remain for at least $90$ days, which surpasses the average lifetime of other bug types.


	\section{Conclusion}\label{sec:conclusion}
We presented in this article a large-scale empirical study that leverages quantitative and qualitative analyses to improve our understanding of mobile code smells. 
The main findings of this study are:
\begin{compactitem}
\item \textbf{Diffuseness}: Android code smells are not introduced and diffused equally.
\emph{No~Low~Memory~Resolver} and \emph{Leaking~Inner~Class} are the most diffuse by affecting $90\,\%$ of activities and inner classes. 

\item \textbf{Releasing pressure}: Releases do not have an impact on the frequency of code smell introductions and removals in open-source Android apps;

\item \textbf{Removal}: $79\,\%$ of code smell instances are removed through the change history.
However, these removals are mostly caused by large source code removals that do not mention refactoring.
Also, only $19\,\%$ of developers who authored these removals confirmed that their actions were intentional refactorings;

\item \textbf{Awareness}: Developers who are aware of Android code smells do not necessarily refactor them. 
The code smell \emph{Init OnDraw} was recognized by $64\,\%$ of the participants, but only $12\,\%$ of them refactored it;

\item \textbf{Refactoring:} Developers who refactored Android code smells were motivated and assisted by built-in code analysis tools and their commitment to code quality.
On the other hand, developers that did not refactor Android code smells doubted their performance impact and the usefulness of their refactoring.
Some developers also preferred to handle performance issues when they arise instead of anticipating them. 
\end{compactitem}
These findings have notable implications on future research agenda:
\begin{compactitem}
	\item We encourage future works to evaluate the impact of releases on mobile code smells in industrial projects and ecosystems.
	This need arises from the remarks of developers about the contrast between the freedom that they have while developing open-source apps and the pressure that they undergo in industrial projects;
	\item Future studies that intend to learn from the change history to build automated refactoring tools cannot rely on code smell removals as learning examples;
	\item To address the questions and doubts raised by developers, we need to reassess the relevance of Android code smells and check the accuracy of their definitions;
	\item We intend to study the relationship between mobile code smells and performance bottlenecks to understand and assess their impact on performance.
\end{compactitem}
Besides, based on our findings, we encourage tool makers to:
\begin{compactitem}
	\item Build profilers that can help developers in spotting performance bottlenecks and identifying their root causes rapidly;
	\item Build and integrate static analysers with more code smell coverage.
	This is beneficial as we observed the impact of such tools on developer awareness and actions;
	\item Build tools that propose automated refactoring of mobile code smells.
	Such tools are crucial for developers who are reluctant toward refactoring by fear of introducing further issues.
\end{compactitem}
This study also provides a comprehensive replication package~\cite{CompanionArtifacts}, which includes the tools, datasets, and results.
	
	\bibliographystyle{elsarticle-harv}
	\bibliography{references} 
	
\end{document}